\documentclass[reprint,amsmath,amssymb,aps,showkeys,floatfix]{revtex4-2}
\usepackage{epsfig,amsfonts,amssymb,amsmath,amsthm}
\usepackage{color}
\usepackage{array}
\usepackage{lineno}
\usepackage{multirow}
\usepackage{graphicx}
\usepackage[caption=false]{subfig}
\usepackage{dcolumn}
\usepackage{bm}

\renewcommand{\epsilon}{\varepsilon}

\begin{document}

\preprint{APS/123-QED}

\title{Robustness of behaviourally-induced oscillations in epidemic models under a low rate of imported cases}

\author{David Juher}
 \email{david.juher@udg.edu}
\author{David Rojas}
 \email{david.rojas@udg.edu}
\author{Joan Salda\~na}
 \email{joan.saldana@udg.edu}
\affiliation{Departament d'Inform\`{a}tica, Matem\`{a}tica Aplicada i Estad\'istica, Universitat de Girona, Catalonia, Spain
}

\date{\today}

\begin{abstract}
This paper is concerned with the robustness of the sustained oscillations predicted by an epidemic ODE model defined on contact networks. The model incorporates the spread of awareness among individuals and, moreover, a small inflow of imported cases. These cases prevent stochastic extinctions when we simulate the epidemics and, hence, they allow to check whether the average dynamics for the fraction of infected individuals are accurately predicted by the ODE model. Stochastic simulations confirm the existence of sustained oscillations for different types of random networks, with a sharp transition from a non-oscillatory asymptotic regime to a periodic one as the alerting rate of susceptible individuals increases from very small values. This abrupt transition to periodic epidemics of high amplitude is quite accurately predicted by the Hopf-bifurcation curve computed from the ODE model using the alerting rate
and the infection transmission rate for aware individuals as tuning parameters.  
\end{abstract}


\maketitle



\section{\label{s1}Introduction}

The importance of the interplay between epidemic spreading and preventive behavioural responses in a globalized world has long been recognized and was specially highlighted after the SARS outbreak of 2003 \cite{Ferguson,FSJ}. The rise of the incidence rate of sexually transmitted diseases (STDs) \cite{CDC2018} and the current resurgence of measles \cite{WHO2012} are also examples of such an interplay. For STDs, increasing high risk sexual behaviour and novel sexual networks are among factors responsible for their re-emergence, whereas vaccine hesitance and distrust in public health intervention programs are among behavioural factors responsible for the rise of diseases like measles.

Risk perception is an important determinant of self-initiated, voluntary protective behaviour \cite{IESBS}. It constitutes the basic ingredient in many epidemic models to encapsulate human behaviour in their formulation \cite{FSJ}. For instance, some extensions of classic deterministic compartmental models include the impact of behaviour on disease transmission by assuming more general incidence rates than the standard one (bilinear). The latter is proportional to the product of the number of susceptible ($S$) and infectious individuals ($I$), $\beta S I$, whereas its generalizations assume a saturation with respect to the number of infectives in order to model a reduction of the contact rate in the presence of a high disease prevalence  \cite{Capasso,Liu, Lu,Ruan}.

Other model extensions take into account awareness transmission among individuals. For example, one of them \cite{Funk10,Kiss10} divides each epidemic compartment ($S$, $I$ and $R$ (recovered/immune)) into two subcompartments of aware and unaware individuals, respectively, and introduces the corresponding transition rates between subcompartments. Other models assume one or more additional compartments consisting of aware ($A$) individuals \cite{Granell, Juher14, Sahneh12a, Sahneh11}. Some of these works consider several awareness levels resulting from the assumption of a degradation in the quality of the information as it is passed from one individual to another \cite{Funk09,JSX}. In all these examples, the effect of preventive behaviour is to modify the values of the epidemic parameters like the probability of infection or the recovery rate.

When the contact network structure of a population is explicitly considered, the effect of behavioural responses can also affect the contact structure itself when modelling social avoidance behaviours. This reduction of the exposure to disease has been modelled by means of preventive disconnection from infectious neighbours \cite{Althouse2014,BJS,Gross06,Gross08,JRS,Llensa,RZ,ZR} or, also, by replacing some infected nodes by healthy ones \cite{Scarpino2016} leading, in both cases, to dynamical networks. Here the assessment of disease prevalence is based on the individual neighbourhood (local contacts), in contrast to homogeneous compartmental models where information about the prevalence is assumed to be globally available \cite{FSJ}.

Under the previous modelling approaches, protective behavioural responses are triggered by the disease prevalence. As long as these responses are based on the global prevalence, one expects the likelihood of epidemic oscillations to be high. Moreover, the linear dependence of the standard incidence rates on the number of infected individuals implies that, when these oscillatory solutions occur, they should pass through low prevalence levels due to the lack of an abrupt switching behaviour. A low prevalence, in turn, will drive the number of aware individuals down as a consequence of a lower perception of the contagion risk, and the cycle repeats again with a new rise in the number of infectives. In fact, several ODE epidemic models with transmission of awareness \cite{JSX, Szabo} or assuming self-initiated, voluntary vaccination \cite{Bauch} exhibit such periodic solutions under some values of the parameters. In particular, the variability in the propensity of aware individuals to further propagate awareness as the  driving mechanism for sustained oscillations was proved in \cite{JSX}. More precisely, it was proved there that, in the absence of demographics, standard incidence rates of infections and awareness do not lead to oscillations even though aware individuals disseminate awareness among susceptible ones, unless a class of individuals with a lower level of awareness (labelled $U$ for unwilling to disseminate) is also assumed.

In this paper we study the robustness of the deterministic oscillations of a model (which we call SAUIS-$\epsilon$) that is an extension of the SAUIS model studied in \cite{JSX}. The reason for this choice is twofold. First, because the only driving mechanism for the existence of periodic solutions in the SAUIS model is the variability in the propensity of alerted individuals to further propagate awareness in the population. Second, because the model does not consider recruitment of susceptible individuals in terms of newborns or in terms of a prevalence-dependent recruitment of them into a core group, which turns out to be an essential requirement for having sustained oscillations in many epidemic models with and without vaccination \cite{Bauch, Ruan, Velasco}. 

Another behavioural mechanism also responsible for the occurrence of periodic solutions in epidemic models is preventive rewiring in adaptive contact networks. Such a mechanism aims to minimise the infection risk of susceptible individuals while maintaining them connected to the network \cite{Althouse2014,Gross06,Gross08,Szabo}. So, both approaches try to explain epidemic oscillations on a pure behavioural basis.  

On the other hand, damped epidemic oscillations have been predicted in \cite{Funk09}, where an SIR epidemic model without demography is coupled with a sophisticated mechanism of degradation of the information quality. Such degradation is translated into individuals with different levels of awareness in the population. However, the eventual depletion of susceptible nodes prevents the occurrence of sustained oscillations.

In countries where an infectious disease has been declared eradicated, new cases can still occur, but these will be isolated and will have limited spread within the community as long as vaccination coverage is high enough. These new cases are usually imported either from tourists, foreign workers, etc., or by local individuals that have been infected abroad in regions where the disease is endemic or large outbreaks are taking place. However, if the vaccination coverage in these countries decreases because of lower levels of awareness (for instance, in countries where vaccination is not mandatory), such few cases can mount into major outbreaks. 

In fact, regions that have achieved the WHO eradication status for a given disease can lose it as a consequence of a marked increase in the number of confirmed cases. For measles, a country attains this status when there is no endemic transmission for 12 months in a specific geographic area. For instance, the UK achieved WHO measles eradication status in 2017, based on data from 2014-2016. However, two years later, this status was lost after 991 confirmed cases in England and Wales in 2018, more than three times the number of cases  (284) in 2017 \cite{UK}.  Other European countries that have also lost the WHO eradication status for measles in 2019 are Albania, Greece, and the Czech Republic \cite{Eradication}. Of course, the role of imported cases is also important in new emergent diseases like Covid-19 where vaccination coverage is not present at all. An illustrative example of their importance is given by the data of Covid-19 in Iceland (as of June 3, 2020): 343 cases out of 1806 were reported to be infected abroad \cite{Iceland}. 

Here, we consider the occasional introduction of new cases into the population, at a very small rate $\epsilon$, the rate of imported cases. From a mathematical point of view, these new cases will not destroy the deterministic oscillations obtained in \cite{JSX} from a Hopf-bifurcation as long as $\epsilon$ is low enough and, at the same time, they prevent stochastic epidemic oscillations from extinction, as it happens when periodic solutions reach very low levels of prevalence (see Subsection~\ref{extincio}). 

The main goal is, then, to analyse the SAUIS-$\epsilon$ model and to perform stochastic simulations of epidemics on networks in order to compare the regions of the parameter space where oscillations occur under both (deterministic and stochastic) approaches. Such a comparison will show that a significant part of the original region of the parameter space where deterministic oscillations occur is preserved when stochastic simulations are performed under the presence of a low rate of imported cases. Moreover, we will see that, as expected, a low level of awareness in the population at the moment of their introduction raises the chances of sparking sustained transmission and, hence, of generating new waves of infection. This could be, for instance, the situation in many countries once lockdown restrictions for Covid-19 have been lifted because most of their inhabitants are not immunized yet, as it happened in Singapore with a second wave of Covid-19 infections within its population of foreigner workers \cite{Singapore}.


\section{\label{sec:SAUISnet}The SAUIS-$\epsilon$ model on a general network}
According to the SAUIS model introduced in \cite{JSX}, each node of a network of size $N$ can be in one of the following four states: S (susceptible), A (aware), U (unwilling), and I (infected). Given any node $n$, let us denote the probabilities that $n$ will have the respective states at time $t$ by $S^n(t)$, $A^n(t)$, $U^n(t)$ and $I^n(t)$. Note that $S^n(t)+A^n(t)+U^n(t)+I^n(t)=1$ for all time $t \ge 0$, since each node has to be in one of the four states. Moreover, for the sake of simplicity, we assume along the paper that the alerting rates per link $\alpha^0_a$, $\nu^0_a$, $\alpha^0_i$, the infection transmission rates per link $\beta^0$, $\beta^0_a$, $\beta^0_u$, the decay rates $\delta_a$, $\delta_u$, and the recovery rate $\delta$ are the same for all the nodes. For instance, if at time $t$ a node $n$ is aware and one of its neighbours, $m$, is susceptible, then 
the probability that $n$ successfully alerts $m$ during the time interval $(t,t+\Delta t)$ is $\alpha_a \Delta t + o(\Delta t)$ provided that $\alpha_a \Delta t < 1$. Similarly, the probability that a non-infected node contracts the infection from abroad (imported case) during a time interval of length $\Delta t$ is $\epsilon \Delta t + o(\Delta t)$. Here it follows a summary of all transitions (or reactions) defining the SAUIS-$\epsilon$ model:
\[ \begin{array}{c}
I + S \stackrel{\beta^0}{\longrightarrow} I + I,\  I + A \stackrel{\beta_a^0}{\longrightarrow} I + I,\ I + U \stackrel{\beta_u^0}{\longrightarrow} I + I \\
I + S \stackrel{\alpha_i^0}{\longrightarrow} I + A,\  A + S \stackrel{\alpha_a^0}{\longrightarrow} A + A,\ A + S \stackrel{\nu_a^0}{\longrightarrow} A + U \\
I \stackrel{\delta}{\longrightarrow} S,\ A \stackrel{\delta_a}{\longrightarrow} U,\ U \stackrel{\delta_u}{\longrightarrow} S, \ \{S,A,U\}\stackrel{\epsilon}{\longrightarrow} I
\end{array}
\]

Let us recall that the SAUIS model tries to account for the degradation of information quality among individuals that are aware of the epidemic situation. As well as in the standard SAIS models, the transition $I+S\rightarrow I+A$ represents the creation of a new aware individual that has acquired first-hand information about the epidemic by means of a direct contact. Also, the transition $A+S\rightarrow A+A$ creates new aware individuals that get indirect information from their acquaintances. Such new aware individuals have the same responsiveness as the information disseminators, which is not always the case. That is why the SAUIS model also includes the $A+S\rightarrow A+U$ transition, where U stands for \emph{unwilling to disseminate information}. So, an unwilling individual has a lower level of awareness, in the sense that he or she does not try to convince other people about the risk, having in addition a weaker behavioural response.

The original SAUIS model in~\cite{JSX} contemplates the possibility that an infected individual, after recovering, may become aware with probability $p$, unwilling with probability $q$, as well as susceptible with probability $1-p-q$. Under these two additional transitions, periodic solutions are also possible (cf. Figs.~7 and 8 in~\cite{JSX}) but, since our ultimate goal is to provide evidence of robustness of the oscillatory regime in non-deterministic epidemics and to simplify the analysis, we will assume $p=q=0$ along the paper. For the sake of simplicity, we have omitted them in the previous description of possible transitions.

Let us derive the approximate discrete-time equations for the evolution of these probabilities. An exact (but unfeasible) description would require the probability of the system being in any of the $4^N$ possible states. So, to derive approximate equations we will assume, as usual, that the joint probability for nodes $n$ and $m$ to be respectively in states $X$ and $Y$ is independent of the neighbourhood's configuration of $n$ and $m$. That is, it equals the product of both probabilities. This hypothesis will allow us to close the system without considering higher order terms for the joint probabilities (see \cite{Juher14} for a related discussion for the $SAIS$ model).

Let $\Delta t>0$ be small enough in such a way that, for every occurrence rate $\kappa$ of a single event, the probability for this event to happen in the time interval $(t,t+\Delta t)$ is $\kappa \Delta t + o(\Delta t)$. For $1\le n,m\le N$, let $a^{nm}$ be the $(n,m)$ element of the $N \times N$ adjacency matrix of the contact network, i.e. $a^{nm}=1$, if the nodes $n$ and $m$ are first neighbours, and $a^{nm}=0$ otherwise. 
With these ingredients, we can now write
\begin{widetext}
\begin{align*}
A^n(t+dt)&=A^n(t)\big(1-\delta_a dt-\epsilon dt -\textstyle (1-\prod_m (1-a^{nm}\beta^0_a dt I^m(t)))\big)\\
&\phantom{=}+ \textstyle S^n(t)\big(1-\prod_m (1-a^{nm}\alpha^0_i dt I^m(t))\prod_m (1-a^{nm}\alpha^0_a dtA^m(t))\big).
\end{align*} 
\end{widetext}

The term multiplying $A^n(t)$ corresponds to the event that the node $n$ keeps being aware at time $t+dt$ provided it was aware at time $t$, so it is 1 minus the sum of the probabilities of the three competing events that change the state $A$ to another one: $A\rightarrow U$ with probability $\delta_a dt$, $A\rightarrow I$ with probability $\epsilon dt$, and the event that one or several infected neighbours of $n$ succeed in performing the transition $I+A\rightarrow I+I$ (with probability $\beta^0_a dt$). For simplicity, the probability of this third event is computed as 1 minus the probability that none of such neighbours succeeds. Analogously, the term multiplying $S^n(t)$ accounts for the probability that $n$ is aware at time $t+dt$ provided it was susceptible at time $t$. Now observe that, neglecting terms of order $o(dt)$, the expressions of the form $\prod_m (1-a^{nm}\kappa dtX^m(t))$ read as $1-\kappa dt\sum_m a^{nm}X^m(t)$. This yields
\begin{widetext}
\begin{align*}
A^n(t+dt)&=A^n(t)(1-\delta_a dt-\epsilon dt - \beta^0_adt\textstyle\sum_m a^{nm}I^m(t))\\
&\phantom{=}+ S^n(t)(1-(1-\alpha^0_idt\textstyle\sum_m a^{nm}I^m(t))(1-\alpha^0_adt\textstyle\sum_m a^{nm}A^m(t))) \\
&=A^n(t)(1-\delta_a dt-\epsilon dt - \beta^0_adt\textstyle\sum_m a^{nm}I^m(t))+ S^n(t)(\alpha^0_idt\textstyle\sum_m a^{nm}I^m(t)+\alpha^0_adt\sum_m a^{nm}A^m(t)),
\end{align*} 
\end{widetext}
where in the second equality we have neglected again the terms of order $o(dt)$.
Subtracting $A^{n}(t)$ to both sides of the previous equation, dividing them by $dt$, and letting $dt \to 0$, we obtain the differential equation governing the time evolution for $A^n(t)$. Proceeding along the same lines for the other probabilities, we finally arrive at the following system of $3N$ ODEs:
\begin{widetext}
\begin{eqnarray}
\frac{d A^n(t)}{dt} &=& \sum_{m=1}^N a^{nm} (\alpha^0_a A^m(t) + \alpha^0_i I^m(t))S^n(t)  - \beta^0_a \sum_{m=1}^N a^{nm} A^n(t) I^m(t) - (\delta_a + \epsilon) A^n(t), \nonumber
\\
\frac{d\,U^n(t)}{dt} &=& \delta_a A^n(t) + \nu^0_a \sum_{m=1}^N a^{nm} S^n(t) A^m(t)  - \beta^0_u \sum_{m=1}^N a^{nm} U^n(t) I^m(t)  - (\delta_u + \epsilon) U^n(t), \qquad \label{SAUISnet}
\\
\frac{d I^n(t)}{dt} &=&  \sum_{m=1}^N a^{nm} \left( \beta^0 S^n(t) + \beta^0_a A^n(t) + \beta^0_u U^n(t) \right)I^m(t) - \delta I^n(t)  + \epsilon \left( 1 - I^n(t) \right), \nonumber
\end{eqnarray}
\end{widetext}
where the equation for $S^n(t)$ is omitted because it is redundant (the sum of the nodal probabilities is always equal to 1). 

From the solution of system \eqref{SAUISnet} endowed with an initial condition, we can compute the expected number of aware, unwilling and infectious nodes at time $t$ by summing the corresponding probabilities over the whole network, that is, $N_A(t)=\sum_n A^n(t)$, $N_U(t)=\sum_n U^n(t)$, $N_I(t)=\sum_n I^n(t)$, and $N_S(t)=N-N_A(t)-N_U(t)-N_I(t)$.

Similar approaches to derive a system of equations for the probabilities for a node of being in one of several disease states have been previously introduced for the study of epidemics on networks and have received different names like, for instance, Microscopic Monte Carlo Approach in a discrete-time setting \cite{Gomez}, or N-intertwined model in a continuous-time setting \cite{Mieghem}. An extension of the latter to multilayer networks is given in \cite{Sahneh13}.

\section{The SAUIS-$\epsilon$ model on regular random networks}\label{SAUIS}

To analyse system \eqref{SAUISnet} we start by the simplest case. So, let us consider the model over a random regular network (not necessarily fully connected) of degree $k$. As we will see, in this particular case the solutions of the system \eqref{SAUISnet} of $3N$ ODEs can be identified with the solutions of a much simpler system of three ODEs.

On this sort of networks, every node has the same vulnerability against the disease (in this setting, the degree is the only characteristic that distinguishes one node from another). So, it is reasonable to assume the same initial probabilities of being aware, $A^n(0)=a_0$, unwilling, $U^n(0)=u_0$, and infected, $I^n(0)=i_0$, for all nodes (what we call a \emph{uniform initial condition}). 

Now we focus on \emph{uniform solutions} of system \eqref{SAUISnet}, defined as those solutions $A^n(t)$, $U^n(t)$, $I^n(t)$ that are independent from $n$. So, we can write $A^n(t)=a(t)$, $U^n(t)=u(t)$ and $I^n(t)=i(t)$ for all $1\le n\le N$. Then, the sums in system~\eqref{SAUISnet} reduce to $\sum_m a^{nm} A^m(t) = k \, a(t)$ and $\sum_m a^{nm} I^m(t) = k \, i(t)$ because each node has the same degree $k$. So, the time evolution of these probabilities satisfies the following initial value problem (IVP):
\begin{equation}\label{eqn:SAUIS}
\begin{split}
\frac{da}{dt} & =  \alpha_i \, s \, i + \alpha_a \, s \, a  -  \beta_a a \, i - \delta_a\, a - \epsilon a, \quad \beta_a < \beta, \\
\frac{du}{dt} & =  \delta_a \, a  + \nu_a \, s \, a -  \beta_u u \, i - \delta_u u - \epsilon u, \quad \beta_u < \beta, \\
\frac{di}{dt} & =  (\beta  \, s + \beta_a a + \beta_u u - \delta) i + (1-i)\epsilon,
\end{split}
\end{equation}
$s+a+u+i = 1$, endowed with the initial condition $a(0)=a_0$, $u(0)=u_0$, and $i(0)=i_0$. Here all the alerting and infection transmission rates are per node and not per link, that is, $\alpha_i = k \, \alpha^0_i$, $\alpha_a = k \, \alpha^0_a$, $\nu_a = k \, \nu^0_a$, $\beta = k \beta^0$, $\beta_a = k \beta^0_a$, and $\beta_u = k \beta^0_u$.

Note that an initial condition $a(0)=a_0$, $u(0)=u_0$, $i(0)=i_0$ of system \eqref{eqn:SAUIS} corresponds to the uniform initial condition $A^n(0)=a_0$, $U^n(0)=u_0$, $I^n(0)=i_0$ for $1\le n\le N$ of system \eqref{SAUISnet}. Then, the local existence and uniqueness of solutions for both IVPs implies that a solution for \eqref{eqn:SAUIS} is also a solution (uniform by construction) for \eqref{SAUISnet}. That is, they are equivalent formulations of the epidemic on regular networks (see Lemma 3.1 in \cite{Juher14} for a proof of a similar result for the SAIS model).

On the other hand, notice that $N_A(t)=a(t)N$, $N_U(t)=u(t)N$, and $N_I(t)=i(t)N$. So, instead of thinking of nodal probabilities, we can consider the expected fraction of aware, unwilling and infected individuals as the macroscopic description (state variables) of our system which is more convenient to compare theoretical predictions with the outputs of the stochastic simulations of the epidemics.

As an incidental remark, it turns out that uniform equilibrium solutions for the complete system \eqref{SAUISnet} of $3N$ equations are only possible over regular networks. To see it, set $A^n=A$, $U^n=U$, $I^n=I$ for all $1\le n\le N$ and denote the degree of any node $n$ by $k_n$. At the equilibrium, we have that $0=dA^n(t)/dt=k_n(\alpha^0_a A + \alpha^0_i I)S -\beta^0_ak_nAI-(\delta_a + \epsilon) A$. Considering a pair of different nodes $n,m$, from $dA^n(t)/dt=dA^m(t)/dt$ we easily get that $k_n=k_m$. In consequence, all nodes must have the same degree. 

\subsection{Behaviour of the equilibria}
Similarly to the SAUIS model in~\cite{JSX}, the tetrahedron $\Omega\!:=\{(a,u,i)\in \mathbb{R}^3 : 0\leq a + u + i \leq 1\}$ is positively invariant under the flow of system~\eqref{eqn:SAUIS}. In fact, the vector field on the boundary of $\Omega$ points strictly towards its interior for $\epsilon > 0$. In consequence, there are no equilibria on the boundary of $\Omega$. 

When $\epsilon=0$ the SAUIS model may have three different kinds of equilibria in $\Omega$: the trivial equilibrium $P_1=(0,0,0)$, the disease-free equilibrium $P_2=(a_0^*,u_0^*,0)$ with
\[
a_0^*=\frac{\delta_u\left(1-\frac{\delta_a}{\alpha_a}\right)}{\delta_a\left(1+\frac{\nu_a}{\alpha_a}\right)+\delta_u},\ u_0^*=\frac{\delta_a\left(1-\frac{\delta_a}{\alpha_a}\right)\left(1+\frac{\nu_a}{\alpha_a}\right)}{\delta_a\left(1+\frac{\nu_a}{\alpha_a}\right)+\delta_u},
\]
and endemic equilibria $P_3=(a^*,u^*,i^*)\in\Omega$ with
\[
i^*=1-\left(1-\frac{\beta_a}{\beta}\right)a^*-\left(1-\frac{\beta_u}{\beta}\right)u^*-\frac{\delta}{\beta}>0.
\]
Note that several distinct endemic equilibria may coexist. When $\epsilon>0$ in the SAUIS-$\epsilon$ model, the first two kinds of equilibria cannot exist and all possible equilibria are endemic. In fact, if $\epsilon$ is a small parameter the equilibria of SAUIS-$\epsilon$ can be interpreted as perturbations of the equilibria $P_1$, $P_2$ and $P_3$ of the unperturbed SAUIS model. To study how these equilibria behave when $\epsilon$ is included in the model, we denote by $f_j(a,u,i;\epsilon)$ the right-hand side of the $j$th equation in~\eqref{eqn:SAUIS}. Any equilibrium $\textbf{e}^*(\epsilon)=(a^*(\epsilon),u^*(\epsilon),i^*(\epsilon))$ of the system is implicitly given by the equations $f_j(\textbf{e}^*(\epsilon);\epsilon)=0$, $j=1,2,3$. We can derive implicitly the equations with respect to $\epsilon$ and get 
\begin{widetext}
\[
\frac{\partial f_j}{\partial a}(\textbf{e}^*(\epsilon);\epsilon)\frac{da^*}{d\epsilon}+\frac{\partial f_j}{\partial u}(\textbf{e}^*(\epsilon);\epsilon)\frac{du^*}{d\epsilon}+\frac{\partial f_j}{\partial i}(\textbf{e}^*(\epsilon);\epsilon)\frac{di^*}{d\epsilon}+\frac{\partial f_j}{\partial \epsilon}(\textbf{e}^*(\epsilon);\epsilon)=0,\ j=1,2,3.
\]
\end{widetext}
In the particular case of the equilibrium $\textbf{e}^*(0)=P_1=(0,0,0)$ the previous system for $\epsilon=0$ can be solved and
\begin{align*}
\left.\frac{da^*}{d\epsilon}\right|_{\epsilon=0}&=\frac{\alpha_i}{(\beta-\delta)(\alpha_a-\delta_a)},\\
\left.\frac{du^*}{d\epsilon}\right|_{\epsilon=0}&=\frac{\alpha_i(\delta_a+\nu_a)}{\delta_u(\beta-\delta)(\alpha_a-\delta_a)},\\
\left.\frac{di^*}{d\epsilon}\right|_{\epsilon=0}&=-\frac{1}{\beta-\delta}.
\end{align*}
In order that the trivial equilibrium $P_1$ of the SAUIS model stays inside the biologically feasible region $\Omega$ when perturbed by $\epsilon$, the three previous expressions must be positive. This occurs when $\beta<\delta$ and $\alpha_a<\delta_a$. In any other case, the SAUIS-$\epsilon$ model has no endemic equilibria bifurcating from $P_1$ for $\epsilon>0$ small. In fact, the eigenvalues of $P_1$ for the unperturbed SAUIS model are
\[
\lambda_1(P_1)=\alpha_a-\delta_a,\ \lambda_2(P_1)=-\delta_u,\ \lambda_3(P_1)=\beta-\delta.
\]
In consequence, $P_1$ bifurcates to an endemic equilibrium for the system SAUIS-$\epsilon$ when $P_1$ is hyperbolic stable in the SAUIS model. Moreover, by the hyperbolic property the equilibrium remains stable for $\epsilon>0$ small in the SAUIS-$\epsilon$ model. The basic reproduction numbers $R_0=\beta/\delta$ and $R_0^a:=\alpha_a/\delta_a$ (see \cite{JSX}) provide a clear interpretation of this fact: for the SAUIS-$\epsilon$ model with $\epsilon>0$ small enough, a stable equilibrium with all coordinates positive and small emerges when $R_0<1$ and $R_0^a<1$, corresponding to a non-spreading, dying out SAUIS epidemic. In this case, the small equilibrium values of aware, unwilling and infected are essentially fed by the introduction of new infection cases at a rate $\epsilon$ rather than by the epidemic propagation itself.

When the perturbation is considered from the equilibrium $\textbf{e}^*(0)=P_2=(a_0^*,u_0^*,0)$, the previous equations imply
\[
\left.\frac{di^*}{d\epsilon}\right|_{\epsilon=0}=\frac{-1}{\beta-\delta-(\beta-\beta_a)a_0^*-(\beta-\beta_u)u_0^*}.
\]
Thus the condition such that the perturbation of the disease-free equilibria is inside the region $\Omega$ for $\epsilon>0$ small is
\begin{equation}\label{leave_cond}
\beta-\delta-(\beta-\beta_a)a_0^*-(\beta-\beta_u)u_0^*<0.
\end{equation}
We point out that the expression on the left-hand side of the previous inequality is the same as the expression of the unique eigenvalue of $P_2$ for the unperturbed SAUIS model that may take positive values (see equation (13) in \cite{JSX} and comments surrounding). The other two eigenvalues are either negative or have negative real part. In particular, this expression is negative if $\beta<\delta$ (since $\beta>\beta_a,\beta_u$), meaning that $\left.\frac{di^*}{d\epsilon}\right|_{\epsilon=0}$ is positive and, so, the SAUIS-$\epsilon$ system has an endemic equilibrium bifurcating from $P_2$ for $\epsilon>0$ small. When $\beta>\delta$, the expression can be positive or negative depending on the other parameters. This change can be controlled by taking $\beta_a$ as a bifurcation parameter and so the bifurcation value is
\[
\beta_a^c\!:=\beta-\frac{1}{a_0^*}(\beta-\delta-(\beta-\beta_u)u_0^*),
\]
as shown in \cite{JSX}. In the SAUIS model, the system shows a transcritical bifurcation as $\beta_a$ passes through the bifurcation value and the authors illustrate that this bifurcation may occur in two different directions. That is, by changing the stability of $P_2$, from an stable equilibrium $P_2$ for $\beta_a<\beta_a^c$ a forward stable endemic equilibrium may bifurcate; or from an unstable equilibrium $P_2$ for $\beta_a>\beta_a^c$ a backward unstable endemic equilibrium may bifurcate. In both cases, the disease-free equilibrium is stable if $\beta_a<\beta_a^c$ and unstable otherwise. In consequence, an endemic equilibrium bifurcates from the disease-free equilibrium in the SAUIS-$\epsilon$ when $P_2$ is hyperbolic stable. As before in the case of $P_1$, the stability is preserved for $\epsilon>0$ small because of the hyperbolic property.

The fact that the equilibria enter the region $\Omega$ for $\epsilon>0$ when they are hyperbolic stable is not surprising. Indeed, the vector field of the SAUIS-$\epsilon$ model at the boundary of $\Omega$ points towards its interior for $\epsilon>0$. This would be in contradiction with a hyperbolic unstable equilibrium entering $\Omega$ from the boundary. 

A similar treatment of the effect of immigration of infected individuals on the disease-free equilibrium of a general epidemic model (without awareness), in terms of  the basic reproduction number, is given in~\cite{AlmCon2019}.

Concerning the endemic equilibria $P_3$, by means of the implicit function theorem, we know that the root $\textbf{e}^*(\epsilon)$ of \eqref{eqn:SAUIS} will persist inside $\Omega$ for $\epsilon>0$ small enough under the classical transversal condition as well as its stability. We refer to \cite{GH} for further information on the dynamical techniques used in this section and the forthcoming one.

\subsection{Robustness of the oscillatory regime} \label{presentaciodelhopf}

\begin{figure}
\centering
    \includegraphics[width=\columnwidth]{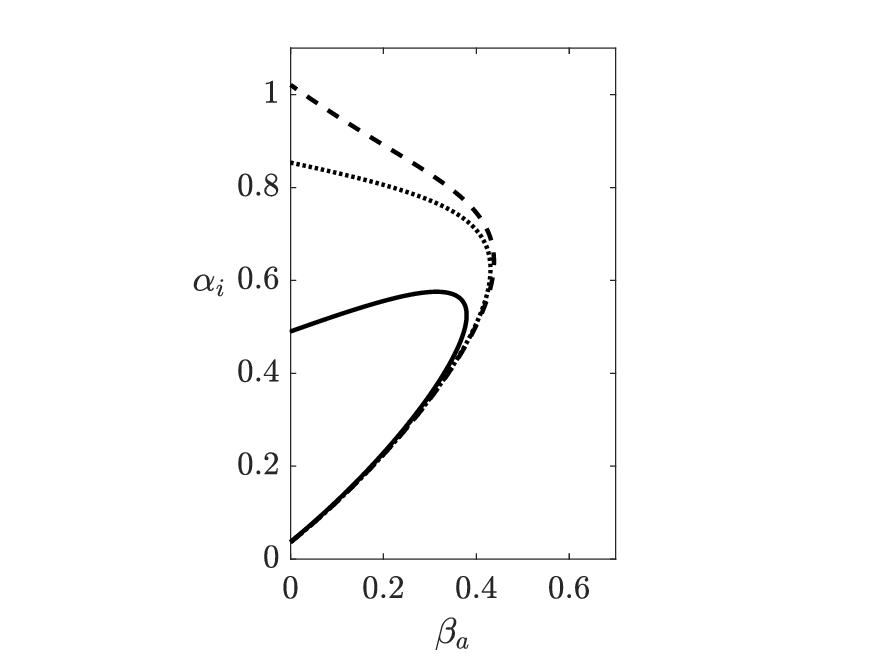}
\caption{Hopf-bifurcation curves in the $(\beta_a,\alpha_i)$ parameter space for $\epsilon=0$ (dashed line), $\epsilon=10^{-5}$ (dotted line) and $\epsilon=10^{-4}$ (solid line). Parameters: $\delta=1$, $\delta_a=0.01$, $\delta_u=0.05$, $\beta=3$, $\beta_u=0.5$, $\alpha_a=0.01$, $\nu_a=1$.}
\label{fig:Hopf}
\end{figure}

The stability of an endemic equilibrium can change under a suitable election of parameters' values. In particular, for $\epsilon=0$, a Hopf-bifurcation curve $H_0$ in the $(\beta_a, \alpha_i)$ parameter space was obtained in \cite{JSX}. Here we also do the analysis for $\epsilon=10^{-5}$ and $\epsilon=10^{-4}$, which are small but still large enough to allow the existence of Hopf-bifurcation curves $H_{\epsilon}$ clearly separated from $H_0$ (see Fig.~\ref{fig:Hopf}). For each value of $\epsilon$, the regime of sustained oscillatory solutions of system \eqref{eqn:SAUIS} lies inside the region of the parameter space limited by $\beta_a=0$ and the corresponding Hopf-bifurcation curve. Outside this region, solutions tend to a stable endemic equilibrium. This behaviour is due to the fact that the real eigenvalue $\lambda_r$ of the Jacobian matrix $J$ of system~\eqref{eqn:SAUIS} at the endemic equilibrium is always negative for any point of the considered region. Actually, $\lambda_r$ is smaller than the real part of the conjugate pair of complex eigenvalues $\lambda_{\pm}$ that constitute, together with $\lambda_r$, the spectrum $\sigma$ of $J$. So, this means that the stability modulus of $J$, namely,
$\max\{\mathrm{Re}(\lambda) \, | \, \lambda \in \sigma(J) \}$, is given by $\mathrm{Re}(\lambda_{\pm})$.

We recall that to compute the Hopf-bifurcation curve in the $(\beta_a, \alpha_i)$ parameter space we need to find the solutions $(\beta_a^*, \alpha_i^*, a^*, u^*, i^*)$ of the system of equations given by the three equilibrium equations of system \eqref{eqn:SAUIS} together with the condition that follows from Theorem 2.1 and Table 1 in \cite{GMS97} which guarantees that the Jacobian matrix $J$ at the endemic equilibrium has a pair of pure imaginary eigenvalues. Precisely, this condition is
\begin{equation} \label{Hopf_cond}
   c_0 - c_1 c_2 = 0 \quad \text{with} \quad c_1 > 0,
\end{equation}
where $c_0=-\det(J)$, $c_1$ is the sum of the principal minors of $J$, and $c_2=-{\rm trace}(J)$.

\begin{figure}
\centering
    \includegraphics[width=\columnwidth]{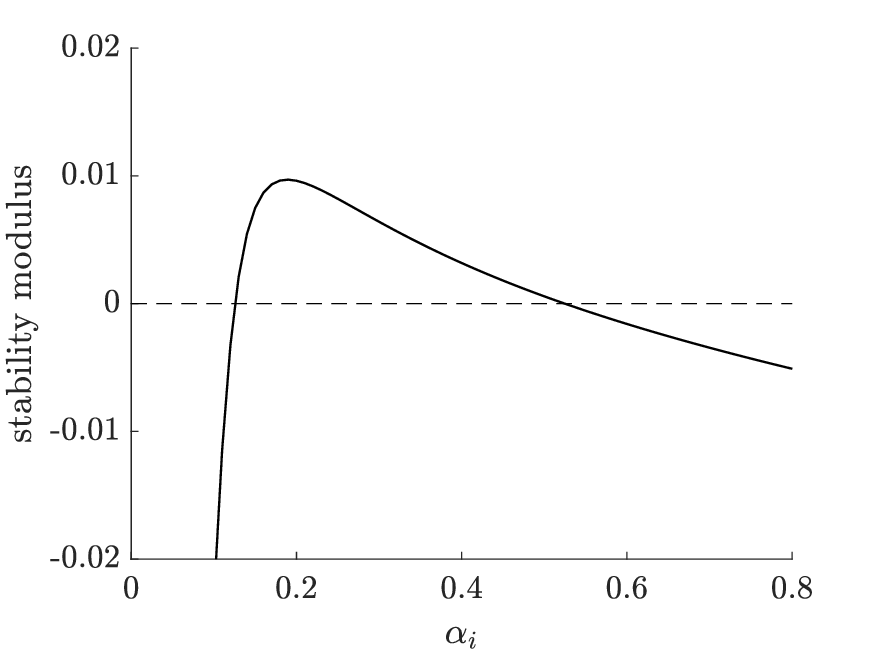}
\caption{Stability modulus of the Jacobian matrix at the endemic equilibrium of~\eqref{eqn:SAUIS} as a function of $\alpha_i$. Parameters: $\delta=1$, $\delta_a=0.01$, $\delta_u=0.05$, $\beta=3$, $\beta_a=0.1$, $\beta_u=0.5$, $\alpha_a=0.01$, $\nu_a=1$, $\epsilon=10^{-4}$.}
\label{fig:partreal}
\end{figure}

From Fig.~\ref{fig:Hopf} we see that, for $\epsilon > 0$, the region limited by $H_{\epsilon}$ is entirely contained in that limited by $H_0$. It follows that the more $\epsilon$ increases, the more the upper part of the original region of the oscillatory regime is reduced, whereas the lower part of this region remains almost the same for the three bifurcation curves. This fact suggests the existence of an abrupt transition from the non-oscillatory regime below the lower branch of the curves to the oscillatory one in the upper side of this branch. This transition is hardly perturbed by the external infection rates $\epsilon$, provided that they are small enough. 

In fact, the qualitative behaviour of the solutions of system~\eqref{eqn:SAUIS} presents significant differences between parameter values near the upper and the lower boundary of the region determined by the Hopf-bifurcation curve. Although all solutions outside the confined zone are foci, a simple study of the eigenvalues of the Jacobian matrix at the endemic equilibrium shows that the transition from focus state (damped oscillations) to a periodic regime is faster through the lower boundary than through the upper one. 

More precisely, it follows that the absolute value of the derivative of the stability modulus that controls the oscillatory motion is much larger for lower values of $\alpha_i$ on the bifurcation curve (lower branch) than for higher ones (upper branch). In Fig.~\ref{fig:partreal} we represent the stability modulus for $\beta_a=0.1$ in terms of $\alpha_i$. This fact becomes crucial to understand the difference of sensibility between the two boundaries in the stochastic Hopf-bifurcation diagrams appearing in next sections. A good example of this fact is the diagram presented in Fig.~\ref{fig:diagrama_hopf_N10000}, where the simulations delimit the lower boundary of the Hopf curve almost identically as the theoretical curve. However, the upper boundary is hardly identified without taking into account the amplitude of the signal.

\section{Stochastic simulations}

\subsection{General simulation setup}\label{sec:Simsetup}
As usual in the setting of continuous-time stochastic simulations, we use the well-known Gillespie algorithm on graphs \cite{Gillespie}. All networks are randomly generated using the configuration model algorithm \cite{brit}. Given a network of size $N$, a combination of model parameters, and an initial condition $a(0),u(0),i(0)$, we run 50 independent simulations, each corresponding to a random distribution of $a(0)N$, $u(0)N$, $i(0)N$, and $(1-a(0)-u(0)-i(0))N$ nodes having respectively the initial states of aware, unwilling, infected, and susceptible. For any experiment, we store the evolution of $a(t)$, $u(t)$, and $i(t)$ as three time series of $2^{12}$ equally-spaced points in the interval $[0,T]$, where $T$ is the maximum running continuous-time of the simulation. All along the paper, the caption of each reported figure obtained by simulation includes the specification of the values of $N$, $a(0)$, $u(0)$, $i(0)$, and $T$.

\subsection{The special case of regular random networks}
It is worth noticing that the classical Gillespie algorithm on graphs is highly time-consuming when executed over a network of about $10^4$ nodes, in such a way that it is not feasible to construct a bifurcation diagram on two parameters, $p_1$ and $p_2$, running 50 experiments for each pair $(p_1,p_2)$, when in addition the number of pairs is of the order of $10^3$. In the particular case of regular random networks, this serious drawback can be overcome by using what we will call the \emph{fast Gillespie algorithm} (FGA in what follows). The FGA crucially relies on the following \emph{mean-field hypothesis} (MFH): on a regular random network of big enough degree, the probability that a neighbour of a node has a given state can be approximated by the fraction of nodes on the entire network having that particular state. Let us see how the MFH can be used to speed up the Gillespie algorithm.

Assume that during an experiment over a given network a susceptible node $n$ gets infected. In this case, the Gillespie algorithm updates the state of $n$ (from $S$ to $I$) and then explores all neighbours of $n$ in order to update the number and type of links to be considered in the next time step. For instance, if a neighbour of $n$ is aware, we lose a link of type $A-S$ with associated weight $\alpha_a^0+\nu_a^0$, and we gain a link of type $I-S$ with associated weight $\beta^0+\alpha_i^0$. This exploration of the neighbours of a node through an adjacency matrix (usually a pointer of pointers), that has to be done at every discrete time step, is one of the main computational loads of the classical Gillespie algorithm on graphs. 

But assume now that the approximation given by the MFH assumption is good enough. Let $N$, $N_S$, and $N_I$ be respectively the total number of nodes, susceptible nodes, and infected nodes in a regular network of degree $k$. If the MFH holds, then the total number of links of type $I-S$ can be simply computed as $kN_I N_S/(N-1)$. Since the infection event $I+S\longrightarrow I+I$ has rate $\beta^0=\beta/k$, the total weight associated to all such events is then $\beta N_IN_S/(N-1)$.
Analogously, the total weight associated to the event $A+S\longrightarrow A+A$, with rate $\alpha_a^0=\alpha_a/k$, would be $\alpha_a N_A N_S/(N-1)$, and so on.

The high speed of the FGA is achieved because the program does not manage any particular network but only three integer variables $N_A$, $N_I$, $N_U$ (absolute numbers of aware, infected, and unwilling nodes respectively), with $N_S=N-N_A-N_I-N_U$. The total weight of all possible events is then
\begin{align*}
R&:=\frac{N_IN_S(\beta+\alpha_i)}{N-1}+\frac{N_AN_S(\alpha_a+\nu_a)}{N-1}+\frac{N_IN_A\beta_a}{N-1}\\
&+\frac{N_IN_U\beta_u}{N-1}+N_I\delta+N_A\delta_a+N_U\delta_u+(N-N_I)\epsilon.
\end{align*}
A particular event as, for instance, $I+S\longrightarrow I+I$ is chosen with probability $(N_IN_S\beta/(N-1))/R$. In this case, we just increase $N_I$ by 1, recompute $R$ according to the previous formula and proceed to the next time step. There is no need to store a particular adjacency matrix and explore the neighbours of any particular node, simply because the MFH assumption allows us to work just with the absolute numbers of aware, infected, and unwilling nodes. Observe that FGA is independent of the degree $k$. In other words, $k$ is not a parameter of the algorithm. The FGA performs statistically exact simulations of Markovian epidemic processes over regular random networks \emph{of high enough degree} (that is, as long as the MFH applies). It is worth mentioning that what we have called FGA can be identified with the original version of the algorithm \cite{Gillespie}, which was aimed at the stochastic simulation of a fully mixed chemically reacting system.

\begin{figure*}
\subfloat[\label{experiments_gil_fast_edo:a}]{
  \includegraphics[width=\columnwidth]{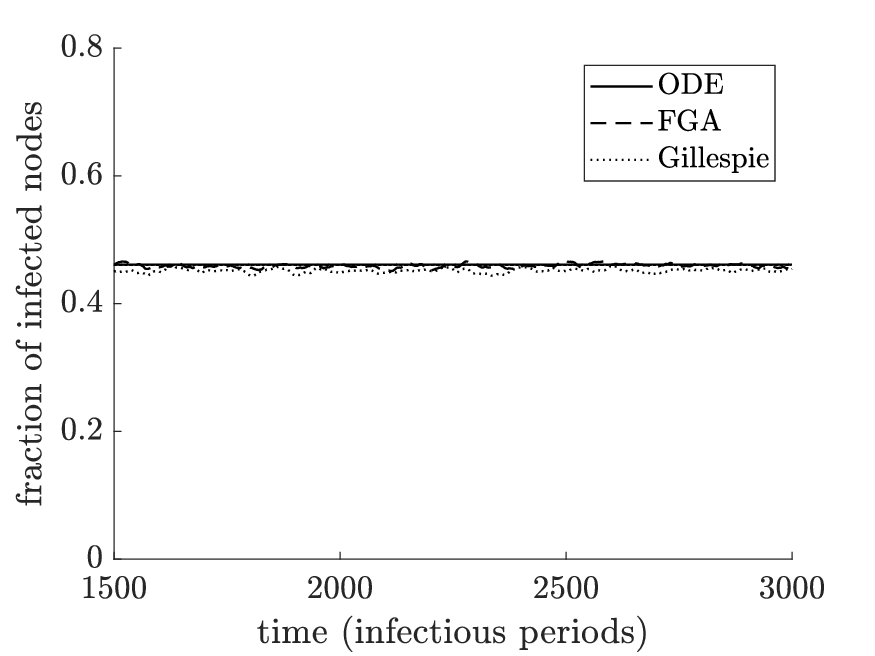}  
}\hfill
\subfloat[\label{experiments_gil_fast_edo:b}]{
  \includegraphics[width=\columnwidth]{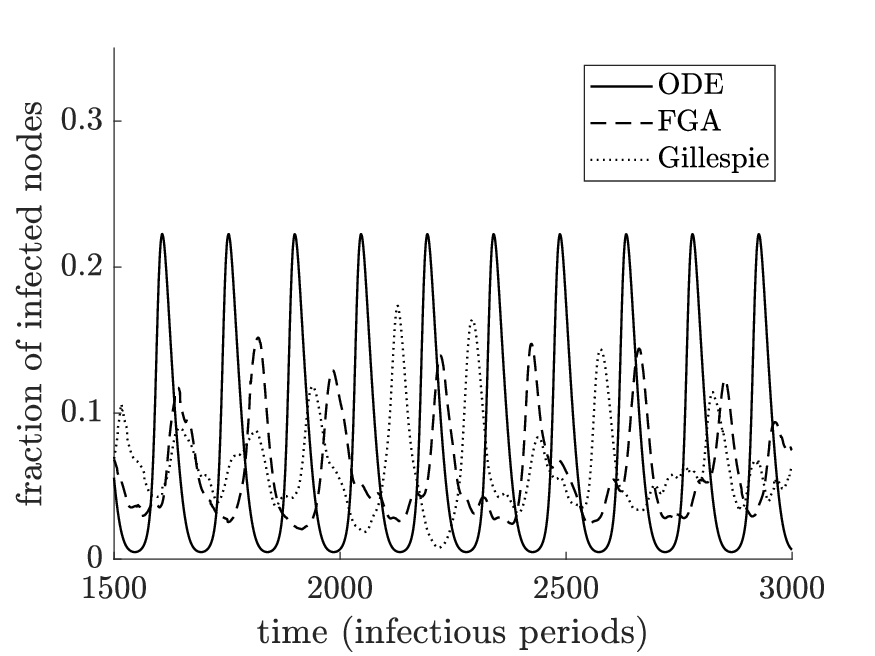}  
}
\caption{Evolution of $i(t)$ for $t\in[1500,3000]$ predicted by system (\ref{eqn:SAUIS}) and 
outputs from Gillespie and FGA (adaptive mean over 50 experiments). Parameters of the simulations: $N=1000$, degree $k=50$ for the Gillespie algorithm, $\delta=1$, $\delta_a=0.01$, $\delta_u=0.05$, $\beta=3$, $\beta_u=0.5$, $\alpha_a=0.01$, $\nu_a=1$, $\epsilon=10^{-4}$, $\beta_a=0.1$, $i(0)=0.1$, $a(0)=u(0)=0.2$. (a) $\alpha_i=0.04$, (b) $\alpha_i=0.14$.}
\label{fig:experiments_gil_fast_edo}
\end{figure*}

Let us see to which extent the outputs of the FGA and the Gillespie algorithm are essentially equivalent. Recall (Subsection~\ref{presentaciodelhopf}) that for $\delta=1$, $\delta_a=0.01$, $\delta_u=0.05$, $\beta=3$, $\beta_u=0.5$, $\alpha_a=0.01$, $\nu_a=1$, and $\epsilon=10^{-4}$, a Hopf-bifurcation curve was obtained in the $(\beta_a, \alpha_i)$-plane. In Fig.~\ref{fig:experiments_gil_fast_edo} we have shown the time evolution of the fraction of infected nodes according to the numerical integration of system (\ref{eqn:SAUIS}), together with the adaptive averaged outputs (see Section~\ref{sec:Significance}) of the Gillespie algorithm over a regular random network of size $N=1000$ and degree 50, and FGA with $N=1000$. When $\alpha_i=0.04$ and $\beta_a=0.1$ (Fig.~\ref{experiments_gil_fast_edo:a}), we are below the Hopf curve and so we have an stable endemic equilibrium. Observe that the three curves are essentially identical. For $\alpha_i=0.14$ and $\beta_a=0.1$ (Fig.~\ref{experiments_gil_fast_edo:b}) we are inside the Hopf curve and we get a stable periodic orbit. In this case, the outputs from Gillespie and FGA \emph{seem} qualitatively equivalent up to stochastic fluctuations. 

\begin{figure*}
\subfloat[\label{comparativaGilFast:a}]{
         \includegraphics[width=\columnwidth]{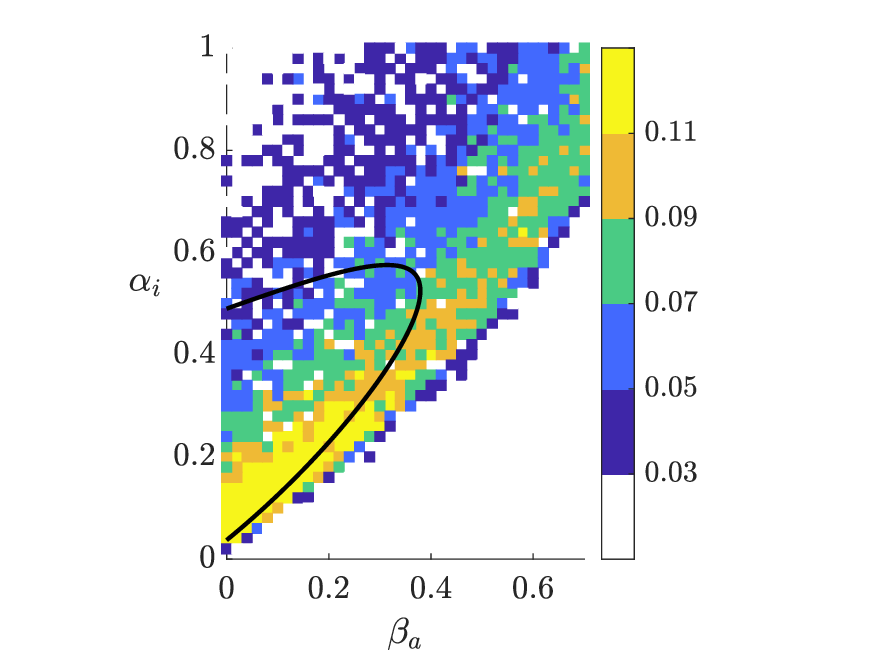}
}\hfill
\subfloat[\label{comparativaGilFast:b}]{
         \includegraphics[width=\columnwidth]{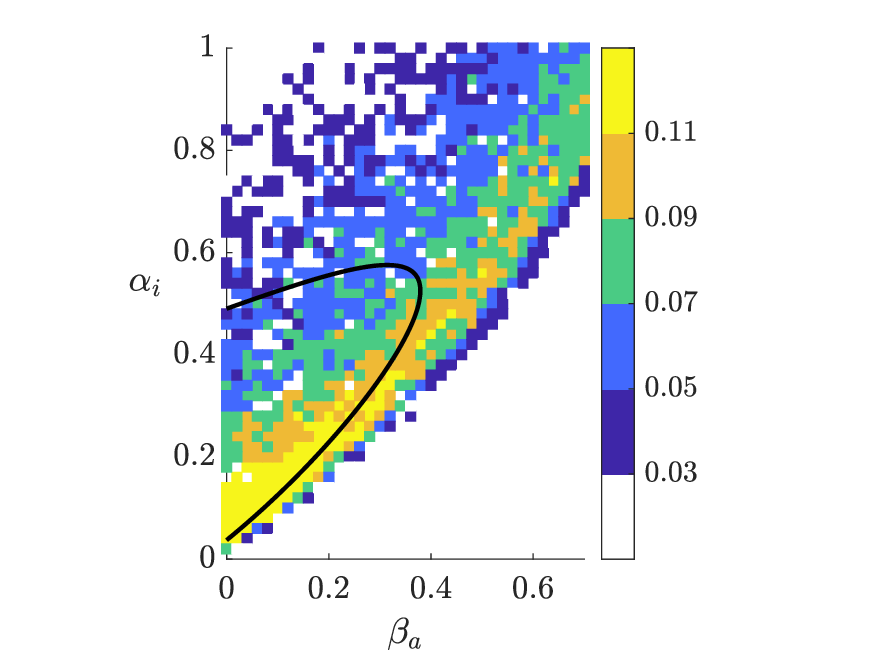}
}
     \caption{Hopf diagram of SAUIS-$\epsilon$ obtained according to the description in Section~\ref{sec:Significance}, using (a) Gillespie algorithm and (b) FGA with $N=1000$ nodes and time $T=3000$. Parameters: degree $k=50$ (for the Gillespie algorithm), $\delta=1$, $\delta_a=0.01$, $\delta_u=0.05$, $\beta=3$, $\beta_u=0.5$, $\alpha_a=0.01$, $\nu_a=1$, $\epsilon=10^{-4}$, $i(0)=0.1$, $a(0)=u(0)=0.2$, 50 experiments for each pair $(\beta_a,\alpha_i)$. The gradient of colours evidences the amplitude of the signal corresponding to the fraction of infected nodes. The black line is the theoretical Hopf-bifurcation curve.}\label{comparativaGilFast}
\end{figure*}

Of course, we should give a precise meaning to the sentence \emph{seem qualitatively equivalent up to stochastic fluctuations}. This is in fact the aim of Section~\ref{sec:Significance}, where we give a detailed explanation about the statistical treatment of the simulation data in order to test the significance of the oscillatory regime. In Fig.~\ref{comparativaGilFast} we show the complete Hopf-bifurcation diagram for the detection of the oscillatory regime in $(\beta_a,\alpha_i) \in [0,0.7]\times[0,1]$ after processing the output data obtained by both the classical Gillespie algorithm over a regular random network of $N=1000$ nodes and degree $50$ (Fig.~\ref{comparativaGilFast:a}) and the FGA with $N=1000$ (Fig.~\ref{comparativaGilFast:b}). We stress that this figure is intended only to compare both algorithms. In particular, we have chosen here $N=1000$ since a higher order for $N$ makes the computation of the Hopf-bifurcation diagrams under the Gillespie algorithm on networks highly costly in time. Observe that the two diagrams are essentially identical, showing that, as expected, the FGA is a good substitute of the Gillespie algorithm on regular graphs even for degrees as small as 50 (over 1000 nodes). 

\subsection{Impact of imported cases in the prevention of stochastic extinctions}
\label{extincio}

Imported cases have an impact on the evolution of an epidemic if they happen when both disease prevalence and awareness level in the population are low. In their absence ($\epsilon=0$), an epidemic becomes eventually extinct because the number of infected nodes becomes extremely low when the level of awareness in the population is very high. In such a situation, it is well known that stochastic extinctions are extremely likely. 

\begin{figure}
\centering
\includegraphics[width=\columnwidth]{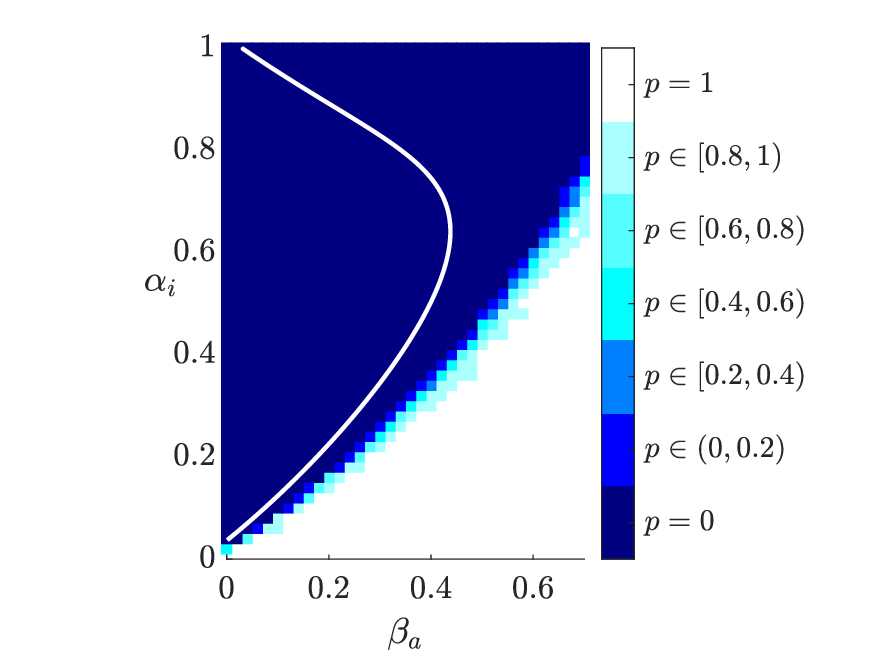}
\caption{Fraction $p$ of stochastic simulations with positive prevalence up to time $T=3000$ for $\epsilon=0$ and $N=10000$ using FGA algorithm. The Hop-bifurcation curve (white curve within the extinction region) is included for a better visualisation of the extinction range. Parameters: $\delta=1$, $\delta_a=0.01$, $\delta_u=0.05$, $\beta=3$, $\beta_u=0.5$, $\alpha_a=0.01$, $\nu_a=1$, $\epsilon=0$, $i(0)=0.1$, $a(0)=u(0)=0.2$, 100 experiments for each pair $(\beta_a,\alpha_i)$.}
\label{fig:extinctions}
\end{figure}

We can observe this fact in Fig.~\ref{fig:extinctions}, which shows that all the region of the $(\beta_a, \alpha_i)$-space where periodic solutions (interior of the Hopf-bifurcation curve) and weakly damped oscillatory solutions are predicted by system \eqref{eqn:SAUIS} lies within the extinction zone (dark region). Note that, for a fixed value of $\alpha_i$, the greater $\beta_a$ is, the higher the disease prevalence at the endemic equilibrium and, hence, the lower the extinction probability of the epidemic. Conversely, for a fixed $\beta_a$, the higher the alerting rate $\alpha_i$ is, the lower the prevalence because aware individuals are more easily created and, hence, the higher the extinction probability is. This is the reason why the non-extinction region corresponds to the lower right part of this figure. 

\begin{figure*}
\subfloat[\label{Input:a}]{
\includegraphics[width=\columnwidth]{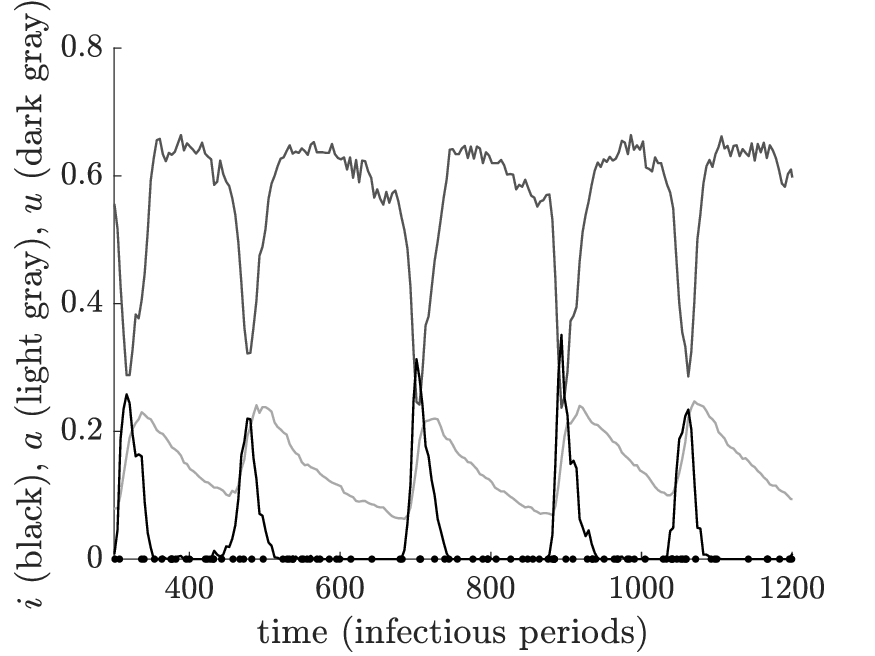}
}\hfill
\subfloat[\label{Input:b}]{
\includegraphics[width=\columnwidth]{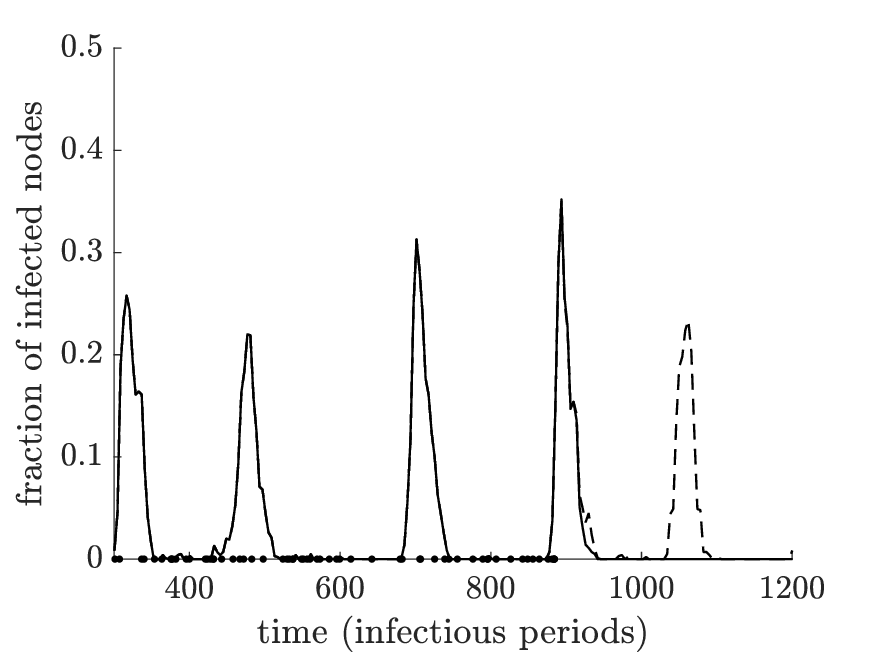}
}
\caption{(a) Time evolution of a stochastic simulation of an epidemic on a random regular network of size $N=1000$ and degree $k=100$ showing the input of imported cases (infections from abroad) for $\epsilon=10^{-4}$ (black dots on the time axis). (b) Fraction of infected nodes in the same simulation until the beginning of the fourth flare-up where $\epsilon=0$ ({\it lockdown}). Dashed line: fraction of infected nodes without lockdown. Parameters: $\delta=1$, $\delta_a=0.01$, $\delta_u=0.05$, $\beta=3$, $\beta_a=0.05$, $\beta_u=0.5$,  $\alpha_i=0.15$, $\alpha_a=0.01$, $\nu_a=1$, $i(0)=0.1$, $a(0)=u(0)=0.2$.}
\label{fig:Input}
\end{figure*}

Fig.~\ref{Input:a} shows one stochastic simulation of an oscillating epidemic on a random regular network of 1000 nodes and, also, the moments where external infections are introduced. Such an oscillatory behaviour is, in fact, predicted by system \eqref{eqn:SAUIS}. However, the extremely low prevalence attained in each cycle makes the stochastic extinction of the disease unavoidable, unless the occasional introduction of imported cases takes place when population awareness is low. The crucial role of such an introduction in maintaining the epidemic is revealed in Fig.~\ref{Input:b}, where the arrival of imported cases has been forced to cease ($\epsilon=0$) just at the beginning of the fourth flare-up. As expected, the fifth flare-up (dashed line) that would appear by keeping $\epsilon=10^{-4}$ now does not happen and the epidemic dies out. The total number of imported cases from $t=0$ to $t=1200$ is 119 in Fig.~\ref{Input:a}, whereas it is equal to 85 in  Fig.~\ref{Input:b} where $\epsilon=0$.

The low number of imported cases in the previous example, about 1 case every 10 infectious periods on average in a population of size 1000, shows that it is not necessary to have a high number of imported cases to prompt the occurrence of important flare-ups in populations whose individuals have a low level of awareness. This situation reminds, for instance, of what happened in New Zealand after the praised management of their first wave of COVID-19, where a reemergence of cases occurred after weeks with no community cases once social restrictions were lifted \cite{Daalder}. 

\subsection{Statistical significance of the oscillatory regime}\label{sec:Significance}

Our approach for the statistical detection of oscillations in the time series has two parts. The first one consists in the generation of three signals from the output of the stochastic simulations, one for each state variable of the model. As stated in Section~\ref{sec:Simsetup}, for each combination of parameters we run $K=50$ independent experiments. Each realisation consists on $2^{12}$ points, of which we eliminate the first half seeking for stationarity of the time series, ending up with $M=2^{11}$ points. In stochastic epidemic models, the usual way to construct a single signal from data is through the average. It is well known that ODE systems are a good approximation of the mean of realisations of stochastic processes in systems with a large number of components (nodes, molecules in reaction systems, etc.). However, although the standard average of trajectories works for endemic equilibria with a high prevalence (see Fig.~\ref{experiments_gil_fast_edo:a}), it is not always a good option when a system exhibits dynamics where some of its components are present in low numbers \cite{Hahl}.  

In particular, the relationship between behaviour and disease we are considering allows the existence of epidemics with periods of very low prevalence once individuals have become aware and adopt efficient protective measures against infection. Such periods are followed by relaxation in the adoption of preventive measures because of a decay in awareness which, in turn, is translated into new epidemic flare-ups. During these time intervals of low prevalence (see Fig.~\ref{Input:a}), the stochastic trajectories of the epidemic show significant random fluctuations which lead to a phase shift in the time series with respect to the deterministic trajectory given by the solution of the ODE model. Consequently, when the average of trajectories is computed, the resulting signal has an important decrease of the amplitude of the oscillations and even a deformation of the periodic component.

To overcome this difficulty, we perform an \emph{adaptive mean} of the signals. The idea is to locally align the signals before computing the average, which acts in favour of preserving periodic motion when the mean is performed. The first part of this method is classic in signal processing. To begin with, we perform a moving average computed over a sliding window of length $21$ centred about each element in the time series. This acts as a low-pass filter to each time series attenuating the signal and omitting extreme frequencies. Then, taking the first time series, say $X_1$, as a sample, we determine the time delay between $X_1$ and each of the other signals. To accomplish this, for each time series $X_k$, $k=1,\dots,K$, we compute the cross-correlation between $X_1$ and $X_k$. The position $\tau_k$ where the maximum is reached, that is
\[
\tau_k\!:=\text{argmax}_{n}((X_1 \star X_k)(n)),\ n=1,\dots,M,
\]
corresponds to the position where the signals are best aligned. For each of the signals $X_k$ we define a new one, $\hat{X}_k$, by applying a circular shift of $\tau_k$ positions to $X_k$. In practice the previous cross-correlation is computed for small delays to ensure that the alignment is local. The adaptive mean we consider is given by
\[
Y \!:= \frac{1}{K}\sum_{k=1}^{K} \hat{X}_k.
\]

\begin{figure*}
\subfloat[\label{periodogrames:a}]{
  \includegraphics[width=\columnwidth]{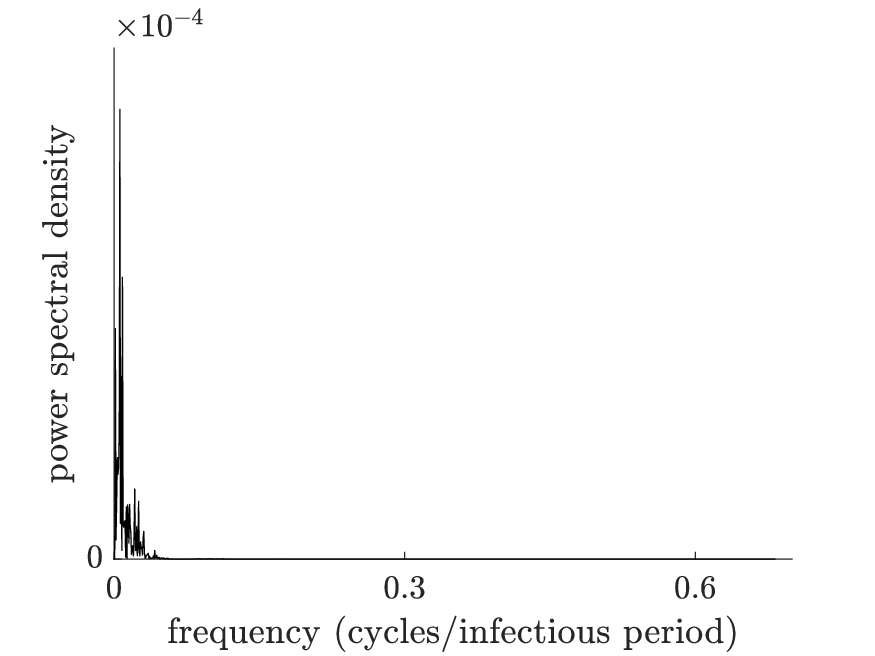}  
}\hfill
\subfloat[\label{periodogrames:b}]{
  \includegraphics[width=\columnwidth]{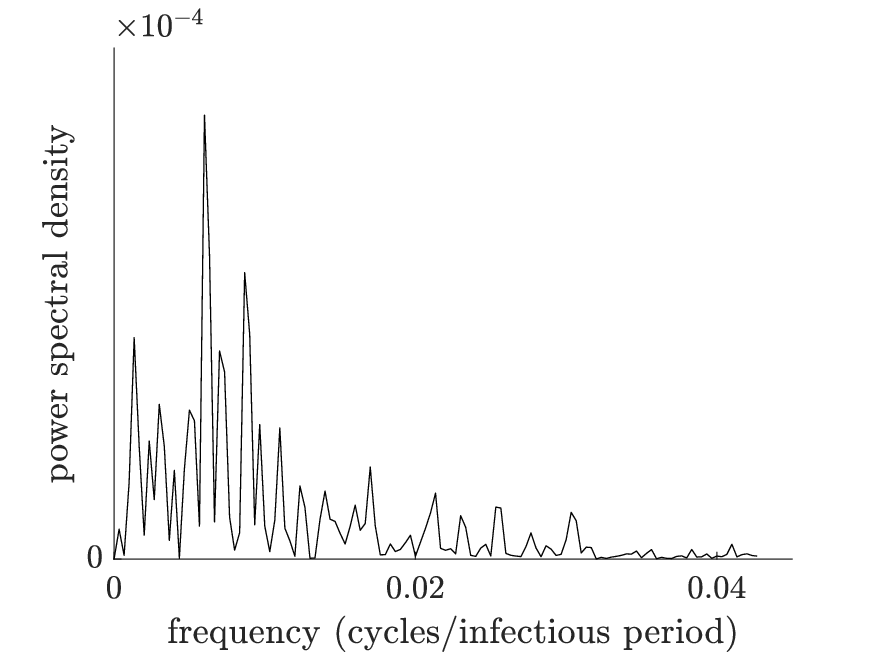}  
}

\subfloat[\label{periodogrames:c}]{
  \includegraphics[width=\columnwidth]{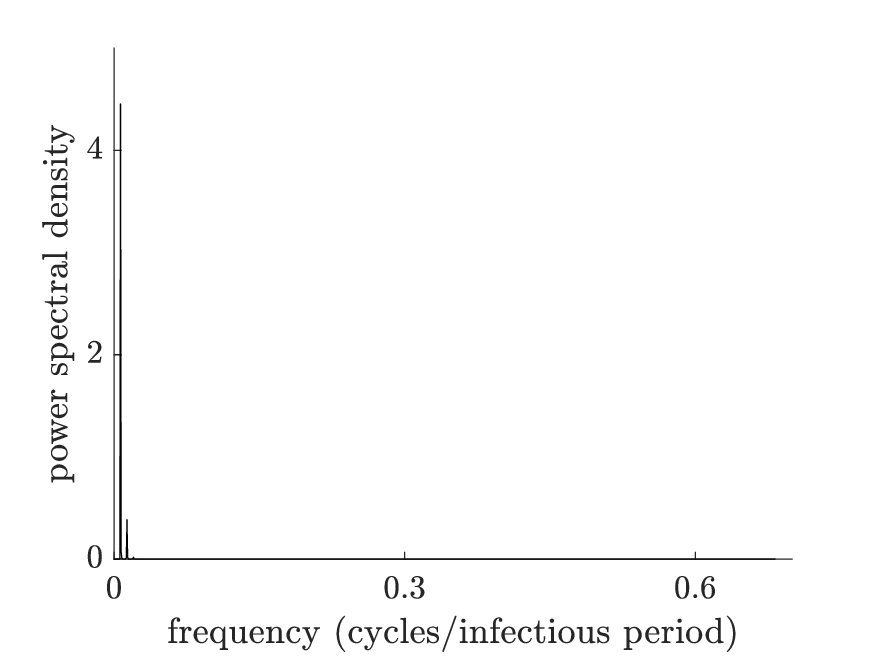}  
}\hfill
\subfloat[\label{periodogrames:d}]{
  \includegraphics[width=\columnwidth]{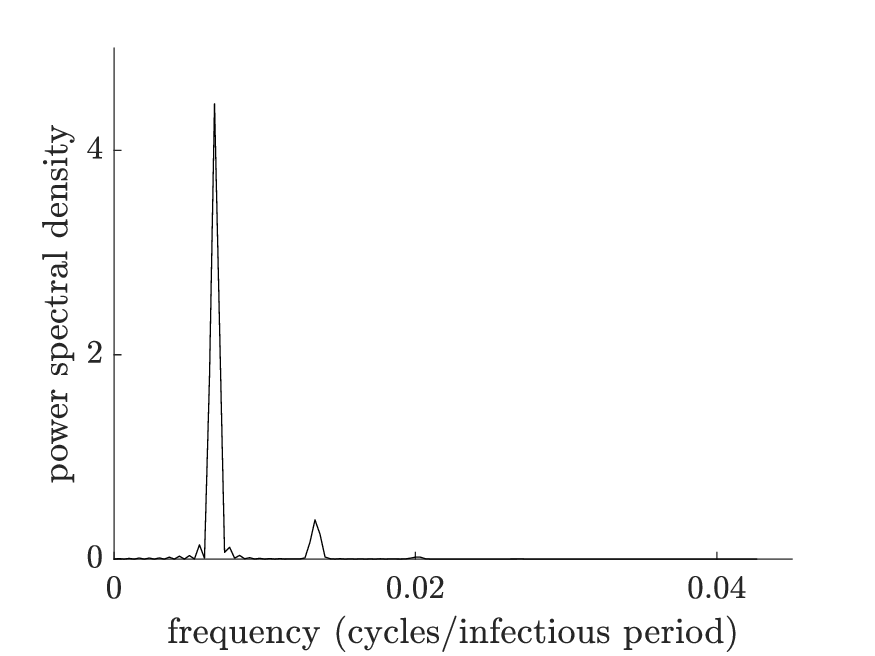}
}
\caption{Periodograms of SAUIS-$\epsilon$ obtained using FGA with $N=10000$ nodes and time $T=3000$ (adaptive mean over $50$ experiments). The number of points used is $M=2^{11}$ for (a) and (c), and $M=2^7$ for (b) and (d), according to the procedure in Section~\ref{sec:Significance}. Parameters: $\delta=1$, $\delta_a=0.01$, $\delta_u=0.05$, $\beta=3$, $\beta_a=0.1$, $\beta_u=0.5$, $\alpha_a=0.01$, $\nu_a=1$, $\epsilon=10^{-4}$, $i(0)=0.1$, $a(0)=u(0)=0.2$. (a) and (b): $\alpha_i=0.04$. (c) and (d): $\alpha_i=0.14$.}
\label{fig:periodogrames}
\end{figure*}

The second part of the detection of oscillations consists in proving the statistical significance of the periodicity of the resulting data $Y$. To do so we define the \emph{periodogram} of the zero-mean time series $\hat{Y}\!:=Y-\overline{Y}$ by
\begin{align*}
P(\omega_0)&=\frac{\Delta t}{M}\left|\sum_{j=1}^{M} \hat{Y}(j) \right|^2,\\
P(\omega_n)&=\frac{2\Delta t}{M}\left|\sum_{j=1}^{M} \hat{Y}(j) e^{-\frac{2\pi n i}{M} (j-1)} \right|^2
\end{align*}
for $n=1,\dots,\frac{M}{2}-1,$ and
\[
P(\omega_{M/2})=\frac{\Delta t}{M}\left|\sum_{j=1}^{M} \hat{Y}(j) e^{-\pi i (j-1)} \right|^2,
\]
with $\omega_n=\frac{n}{M\Delta t}$ and $\Delta t$ being the time-step of the time series. The periodogram is suitable to find hidden periodicities on the data. Indeed, data following a pure random process show a flat periodogram. However, if the time series presents a periodic motion then the frequencies involved are shown as peaks of the periodogram. Fisher's $g$-test \cite{Fis29} is commonly used in this scenario to reject the null hypothesis that the random process is Gaussian white noise against the alternative hypothesis that the series contains a deterministic periodic component of unspecified frequency. The statistic taken into account to reject the null hypothesis is the maximum of the periodogram compared with the sum of all the values of the periodogram. That is,
\[
g=\frac{\max_{n}P(\omega_n)}{\sum_{n=1}^{M/2}P(\omega_n)}.
\]
The previous maximum, as well as the summation, is taken for $n=1,\dots,M/2$. This is so because the term $P(\omega_0)$ contains only information about the mean of the data. This term plays no role in the detection of oscillatory motion and, in fact, vanishes for zero-mean time series. The test is performed by computing the realised value $g^*$ of $g$ from the data and then use the exact distribution of $g$ given by Fisher in \cite{Fis29},
\[
P(g>x)=\sum_{k=1}^b \frac{(-1)^{k-1}\left(\frac{M}{2}\right)!(1-kx)^{\frac{M}{2}-1}}{k!\left(\frac{M}{2}-k\right)!},
\]
where $b$ is the largest integer less than $1/x$. If the probability $P(g>g^*)$ is less than $\alpha$ then the null hypothesis can be rejected at level $\alpha$. For further details on Fisher's test for hidden frequencies we refer to \cite{BroDav91,Pri81,Sch98}.

For the time series resulting from the parameters we are interested in, the direct application of the Fisher's test results positive for all configurations. This is so for two reasons. First, all motions of the ODE system~\eqref{eqn:SAUIS} have an oscillatory component since they correspond to either a focus or a periodic orbit. Second, for the expression of the probability above, it easily follows that the larger the number of points $M$, the smaller the value of the probability. To be more precise in the determination of oscillatory motions, we reduce the points of the signals after the adaptive mean in a ratio $1:2^{4}$. That is, we end up with $M=2^7$ points. This procedure enables to detect the strongest periodic motions in the parameter space, since weaker periodic signals will not pass Fisher's test with fewer points. Fig.~\ref{fig:periodogrames} illustrates it with two parameter configurations. On the left, periodograms with $2^{11}$ points pass Fisher's test for both configurations. On the right, the periodograms of the same averaged signals with $2^7$ points. In this case the configuration corresponding to the periodogram on the top side does not pass Fisher's test, whereas the one on the bottom side still passes the test. Notice that these parameter configurations correspond to the signals of top panels in Fig.~\ref{fig:grafiques}. It is worth to mention that the frequencies avoided by the reduction of points are much larger than the frequencies of the data as can be seen in Fig.~\ref{fig:periodogrames} and information about the periodicity of the signal is not lost.

For those parameters passing the Fisher's $g$-test with a $p$-value less than or equal to $\alpha=0.01$ we consider the corresponding peak frequency of the time series. The bifurcation diagram shows those parameters with an estimated frequency larger than the minimum observable frequency $f_{min}=1/1500$. A gradient of colours illustrates the amplitude of the averaged signal corresponding to infected nodes. In Fig.~\ref{fig:diagrama_hopf_N10000} we have shown the corresponding bifurcation diagram for the outputs of the SAUIS-$\epsilon$ model using the FGA with $N=10000$. The threshold below which oscillations are considered to be internal random fluctuations (noise) of the system is given by the inverse of the square root of the number of nodes, namely, $0.01$ (and $0.03$ when $N=1000$ as in Figs.~\ref{comparativaGilFast} and \ref{ER_Exp}). This value comes from the so-called linear noise approximation used in modelling of chemical reaction kinetics \cite{Grima, VK}. Such an approximation assumes that, in a system formed by different ``chemical species" (nodes in different states in our context), the standard deviation of random fluctuations about the mean number of molecules of these chemical species scales as the square root of the size of the system (the number $N$ of nodes in our network). So, dividing these mean numbers by $N$, it follows  that, under this approximation, the standard deviation of the random fluctuations of the fractions of $A$, $U$, and $I$ nodes is proportional to $N^{-1/2}$.

\begin{figure}
    \centering
    \includegraphics[width=\columnwidth]{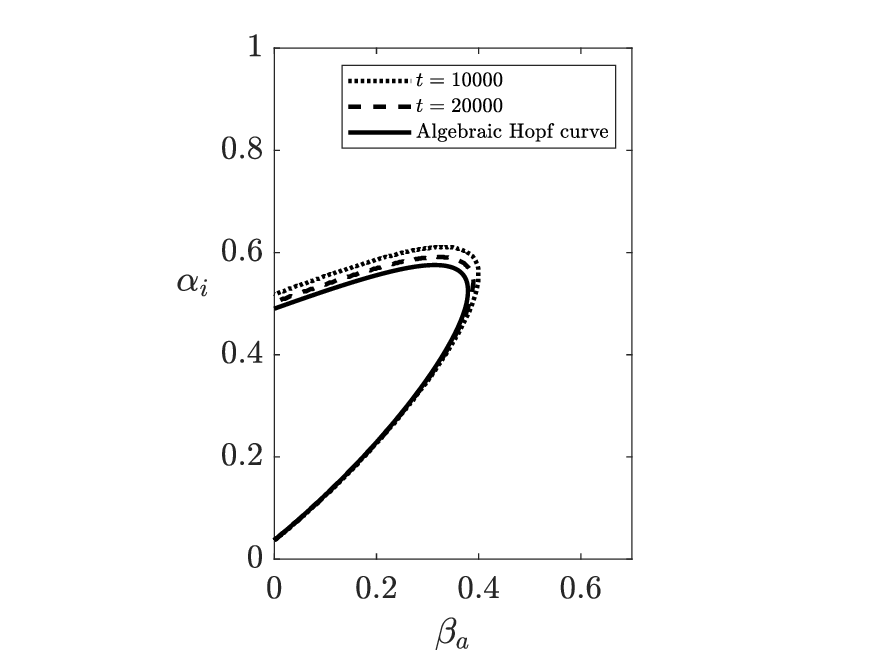}
    \caption{Comparison of the algebraic Hopf-bifurcation curve for regular random networks with two approximations obtained integrating system \eqref{SAUISnet} up to two different times and using for $CV=0.1\%$ to discriminate between periodic and non-periodic solutions.}\label{comparativaHopf}
\end{figure}

\subsection{Oscillations in other network architectures}
Another aspect of the robustness of the oscillations is their likelihood when other network topologies are considered. Are they still present when the epidemic spreads on more heterogeneous networks? If this is the case, are they observed in the solutions of system \eqref{SAUISnet}? To answer these questions, we have used the configuration model algorithm to generate a Poisson network with mean degree 50 and an exponential network with mean degree 50 and minimum degree 25. The first degree distribution corresponds to the well-known Erd\"os-R\'enyi random graphs, whereas exponential degree distributions have been observed in empirical contact networks \cite{Bansal} and have a much higher variability.

\begin{figure}
    \centering
    \includegraphics[width=\columnwidth]{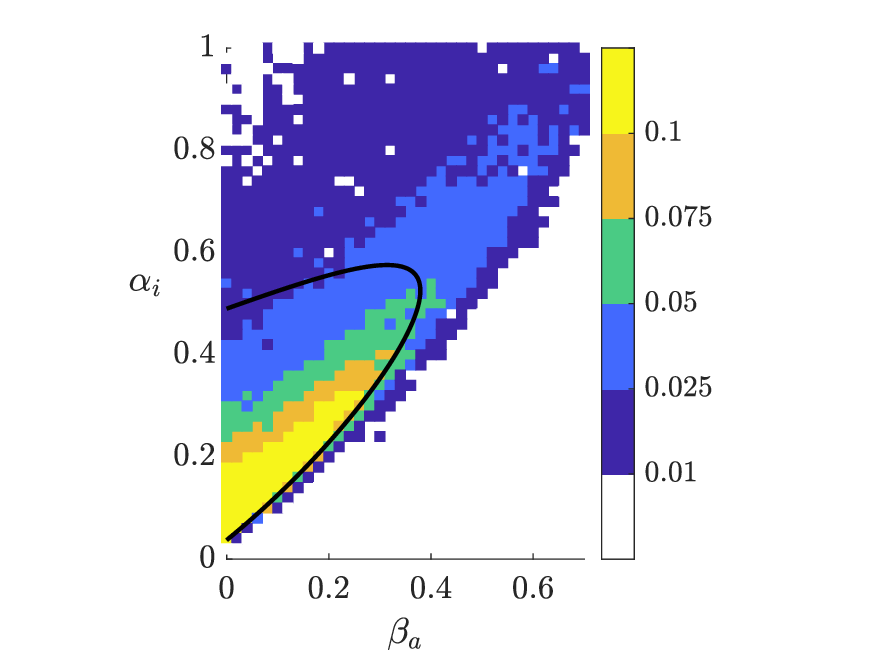}
    \caption{Hopf diagram obtained according to the description in Section~\ref{sec:Significance} of the FGA with $N=10000$ nodes and time $T=3000$. The gradient of colours evidences the amplitude of the averaged signal of the fraction of infected nodes. The black line is the algebraic Hopf-bifurcation curve for $\epsilon=10^{-4}$.}
    \label{fig:diagrama_hopf_N10000}
\end{figure}

Unfortunately, in this case it is not possible to reduce \eqref{SAUISnet} to a simpler system, as it was done in Subsection~\ref{presentaciodelhopf} for regular random networks. So, the algebraic approach used there for computing Hopf-bifurcation curves is no longer feasible. Instead, using $\beta_a$ and $\alpha_i$ as tuning parameters (with increments of size 0.002 or even smaller when approaching the turning point of the curve), we have obtained an approximation to these curves by numerically integrating system \eqref{SAUISnet} for $N=1000$ (i.e. the full system of 3000 equations) until a long enough time $T_2$. Then, for each pair $(\beta_a, \alpha_i)$, we computed the coefficient of variation ($CV$) of the fraction of aware, unwilling and infected nodes from the solution $A^i(t)$, $U^i(t)$, and $I^i(t)$ of \eqref{SAUISnet} for the last 1000 units of time ($T_2-T_1 = 1000$). Precisely, for the fraction $x(t)=\sum_i x^i(t)/N$ where $x^i(t) = A^i(t)$, $U^i(t)$, $I^i(t)$, respectively, we compute $CV_{x(t)} = \sigma_{x(t)}/\bar{x} \times 100$ with
\[
\bar{x}= \frac{1}{T_2-T_1} \int_{T_1}^{T_2} x(t) \, dx
\]
and
\[
\sigma^2_{x(t)} = \frac{1}{T_2-T_1} \int_{T_1}^{T_2} \left( x(t)-\bar{x} \right)^2 dx.
\]

Finally, an approximate Hopf-bifurcation curve is obtained by using the criterion that, for a given pair $(\beta_a, \alpha_i)$, a solution is not periodic if the corresponding $CV \le 0.1\%$ for the three fractions of nodes. Note that, close to the Hopf-bifurcation, solutions classified as periodic when we integrate the system up to a given time can become non-periodic when longer times are considered. This is particularly relevant when trying to delimit the upper branch of the bifurcation curve. Here the transition from periodic solutions to very weakly damped solutions is hardly noticed because of its flatness, which was already observed when computing the algebraic curve for the regular case (see Fig.~\ref{fig:partreal}). 

\begin{figure*}
\subfloat[\label{ER_Exp:a}]{
         \includegraphics[width=\columnwidth]{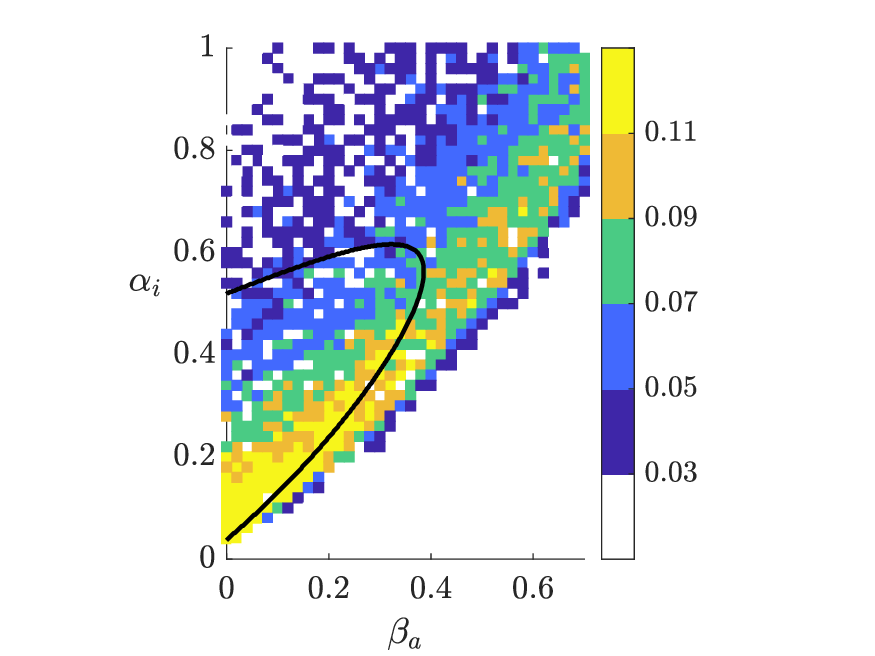}
}\hfill
\subfloat[\label{ER_Exp:b}]{
         \includegraphics[width=\columnwidth]{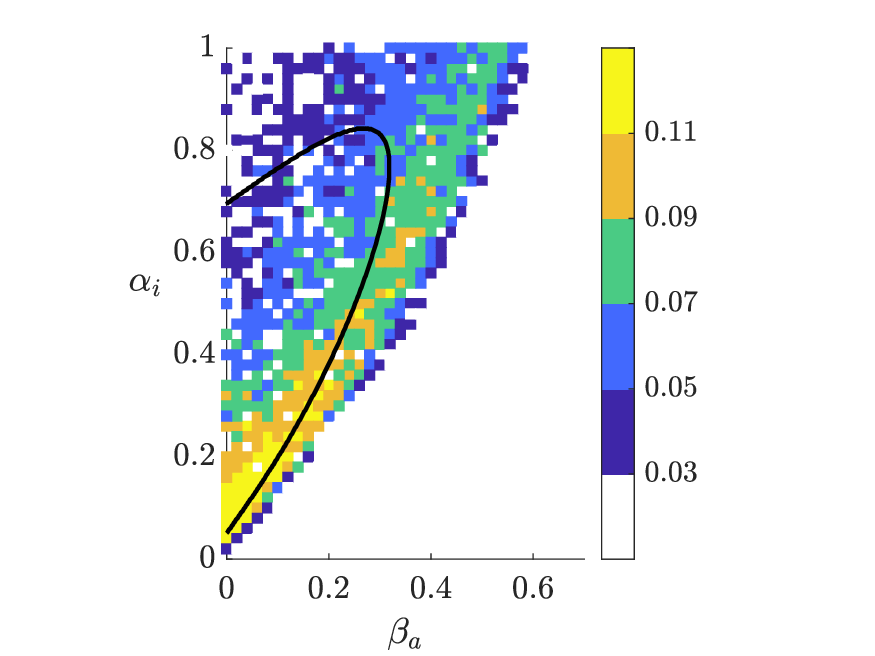}
}
      \caption{Hopf diagram of SAUIS-$\epsilon$ obtained according to the description in Section~\ref{sec:Significance} with $N=1000$ nodes and time $T=3000$, in two different network architectures of mean degree $50$: (a) Poisson and (b) Exponential. Parameters: $\delta=1$, $\delta_a=0.01$, $\delta_u=0.05$, $\beta=3$, $\beta_u=0.5$, $\alpha_a=0.01$, $\nu_a=1$, $\epsilon=10^{-4}$, $i(0)=0.1$, $a(0)=u(0)=0.2$, 50 experiments for each pair $(\beta_a,\alpha_i)$. The gradient of colours evidences the amplitude of the signal corresponding to the fraction of infected nodes. The black line is the theoretical Hopf-bifurcation curve.}
      \label{ER_Exp}
\end{figure*}

To calibrate our criterion with respect to the integration time, we compared the algebraic Hopf-bifurcation curve for system \eqref{eqn:SAUIS} with those obtained from system \eqref{SAUISnet} for a regular random network of degree 50. Notice that system \eqref{eqn:SAUIS} and, hence, the corresponding algebraic Hopf-bifurcation curve, do not depend on the degree, while system \eqref{SAUISnet} does depend on it. Fig.~\ref{comparativaHopf} shows that, as expected, increasing the integration time from $T_2=10000$ to $T_2=20000$ leads to a better approximation to the whole algebraic curve (solid line). This figure also shows that, for $T_2=20000$, the disagreement with respect to the algebraic curve is only perceptible in its upper branch. Therefore, the (approximate) Hopf-bifurcation curves for Poisson and exponential networks were constructed from the solutions of system \eqref{SAUISnet} using an integration time $T_2=20000$ and a threshold value of $CV$ equal to 0.1\% for the fraction of the three types of non-susceptible nodes.

\subsection{Simulation results and discussion}\label{sec:Discussion}
The values of the parameters used in the simulations are taken from \cite{JSX} (except for the rate of imported cases $\epsilon$) and reflect what we consider it is a natural scenario, although they are not intended to model any particular disease. First, aware (and unwilling) individuals are affected by lower transmission rates because of the adoption of preventive measures. Second, the mean infectious period is much shorter than the mean duration of awareness. Finally, it is assumed that is more difficult for an aware individual to convince a susceptible one to become aware than to convince him to become simply unwilling, i.e., to adopt preventive measures but without willingness to convince others about the risk of infection. The values of the awareness and unwillingness decay rates lead to oscillations whose period is about 1.3 times the sum of the mean duration of the awareness period ($1/\delta_a=100$) and the mean duration of the unwillingness period ($1/\delta_u=20$). 

\begin{figure*}
\subfloat[\label{grafiques:a}]{
  \includegraphics[width=\columnwidth]{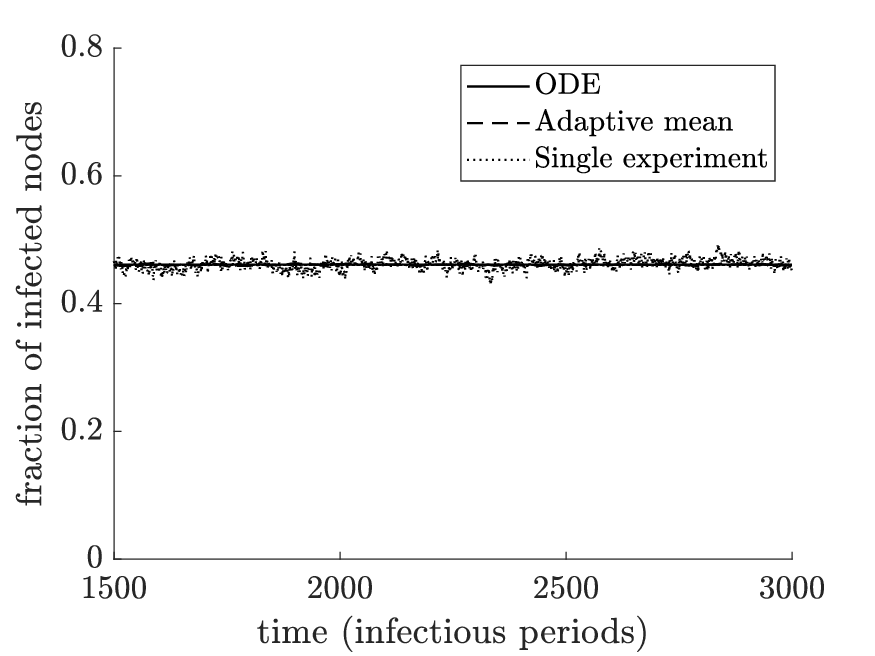}  
}\hfill
\subfloat[\label{grafiques:b}]{
  \includegraphics[width=\columnwidth]{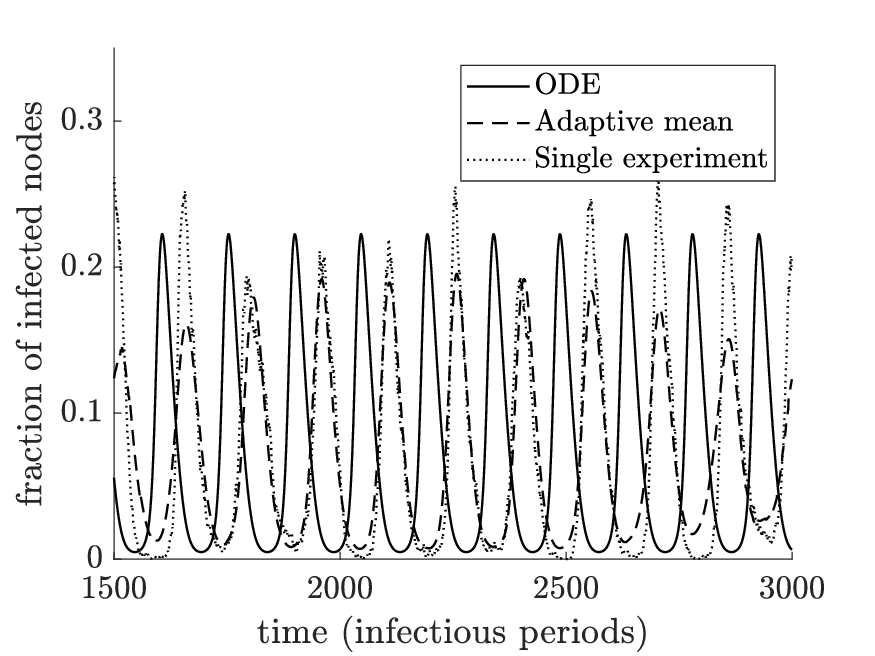}  
}

\subfloat[\label{grafiques:c}]{
  \includegraphics[width=\columnwidth]{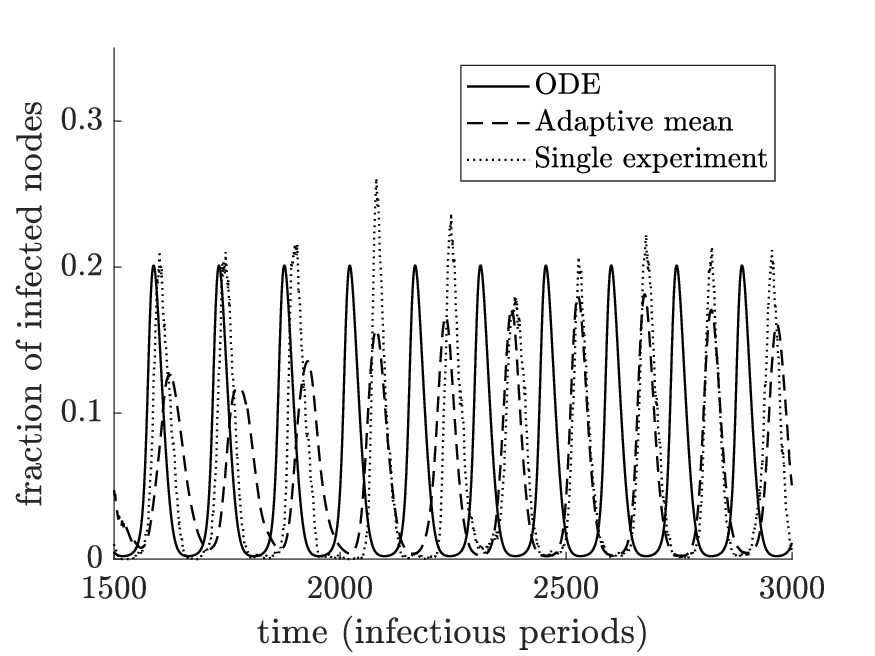}  
}\hfill
\subfloat[\label{grafiques:d}]{
  \includegraphics[width=\columnwidth]{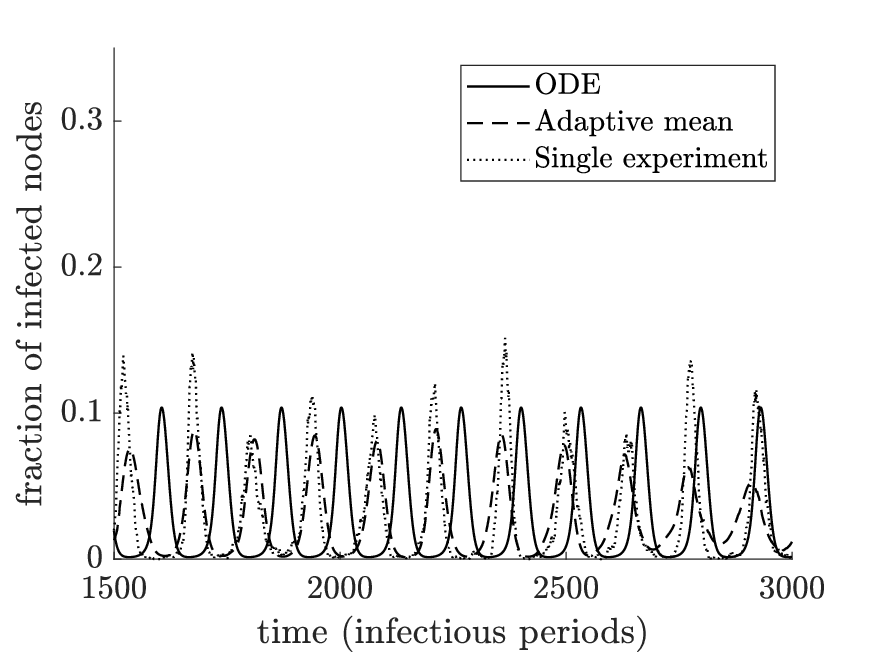}  
}
\caption{Evolution of $i(t)$ for $t\in[1500,3000]$ predicted by system (\ref{eqn:SAUIS}) (solid line), adaptive mean of outputs from FGA (over 50 experiments, dashed line) and a single experiment from FGA (dotted line). Parameters of the simulations: $N=10000$, $\delta=1$, $\delta_a=0.01$, $\delta_u=0.05$, $\beta=3$, $\beta_a=0.1$, $\beta_u=0.5$, $\alpha_a=0.01$, $\nu_a=1$, $\epsilon=10^{-4}$, $i(0)=0.1$, $a(0)=u(0)=0.2$. (a) $\alpha_i=0.04$, (b) $\alpha_i=0.14$, (c) $\alpha_i=0.16$, (d) $\alpha_i=0.26$.}
\label{fig:grafiques}
\end{figure*}

The procedure of the adaptive mean of the signals from the stochastic simulations gives an excellent reconstruction of the oscillatory behaviour of single trajectories. In Fig.~\ref{fig:grafiques}, we show the comparison of a solution of system \eqref{eqn:SAUIS}, the adaptive mean of the 50 realisations of an epidemic on a regular random network of size $N=10000$ generated by means of the FGA, and the trajectory used as the reference for the alignment of the signals. Note that this alignment preserves the periodicity of the original signals (trajectories) but, at the same time, it reduces their amplitudes. With this respect, it is worth recalling that the amplitudes shown in the Hopf diagrams do not correspond to stochastic simulations of single epidemics, but to the adaptive means of 50 stochastic trajectories for each pair $(\beta_a, \alpha_i)$ of parameter values (see Sect.~\ref{sec:Significance} for details). 

Hopf-bifurcation diagrams show that the abrupt transition between a non-oscillatory regime and an oscillatory one given by the lower branch of the Hopf-bifurcation curve is quite accurately captured in all the settings we have considered. The robustness of this transition is, in fact, already observed when this curve is computed for different rates of imported cases (see Fig.~\ref{fig:Hopf}).

The best agreement is achieved for regular random networks of size $N=10000$. Fig.~\ref{fig:diagrama_hopf_N10000} shows that the region with periodic solutions with amplitudes larger than 0.05 falls neatly within the limits predicted by the Hopf-bifurcation curve. Comparing this figure with the panels in Fig.~\ref{comparativaGilFast} obtained for $N=1000$, one can see that the agreement of the observed region of periodic solutions with high amplitudes (yellow and orange squares) with the predicted one clearly increases with the number of nodes. Moreover, we recall that all the simulations have been carried out until time $T=3000$ which can be not enough for weakly damped oscillations to disappear.

A similar agreement with the predictions as the one in Fig.~\ref{comparativaGilFast} is also observed in networks with different topologies but with the same number of nodes, $N=1000$ (see Fig.~\ref{ER_Exp}). Moreover, we can observe that the differences between the Hopf-bifurcation curves when the degree distributions have different variability are remarkable. For regular random networks and Poisson networks, these differences are small, i.e.~both types of networks lead to similar (but not equal) Hopf-bifurcation curves. However, the resulting curve for an exponential network is clearly different, with a turning point at a higher value of $\alpha_i$ and lower values of $\beta_a$ (see Fig.~\ref{ER_Exp:b}). Interestingly, the region of the $(\beta_a, \alpha_i)$ parameter space corresponding to periodic solutions with large amplitudes (yellow squares and most of the orange ones) clearly falls inside the region limited by the Hopf-bifurcation curve.   

All in all, stochastic simulations confirm the predictions of the ODE model showing that the proposed mechanism for prompting awareness is able to generate recurrent epidemic cycles on different network architectures, including the one with an exponential degree distribution which has been claimed to describe contact patterns in real populations. Interestingly, in all cases the predicted region of the parameter space for the oscillatory regime satisfy that $\beta_a < \beta_u$, a natural condition according to the higher level of alertness assumed for aware individuals with respect to the unwilling ones. Moreover, these regions always contain those averaged stochastic trajectories with the highest amplitudes.

\section{Conclusions}
The interplay between human behaviour and epidemic spread has been considered in many papers during the last 15 years \cite{FSJ,Verelst}. Most of them have as the main goal to elucidate the dependence of the basic reproduction number on the different behavioural responses as, for instance, social distancing, rewiring of connections, awareness, etc.~\cite{Funk10,Juher14,Kiss10,Llensa,Sahneh12a,ZR}. Few of them address the existence of oscillating epidemics  \cite{Bauch,Gross06,Gross08,Szabo,Velasco} and, as far as we know, only in \cite{JSX} sustained oscillations arise uniquely from the interaction between awareness dissemination and epidemic spread, i.e.~without the need of any recruitment of new (susceptible) individuals.

In this paper we have challenged the existence of oscillations predicted by the SAUIS model introduced in \cite{JSX} by means of stochastic simulations on different network topologies. The model has been formulated on networks and considers the existence of a very small inflow of imported cases. These cases are essential to keep the epidemic going on because disease prevalence attains very low levels when there is a high degree of awareness in an oscillating epidemic. This is not a problem at all in a deterministic framework, but it leads to an unavoidable stochastic epidemic extinction in all parameter combinations where sustained oscillations are present.      

Our simulations show that the presence of a small number of imported cases allows, when the disease prevalence is low and the number of aware and unwilling individuals decreases, an oscillatory behaviour of the stochastic epidemics which closely resembles the one predicted by the deterministic counterpart of the model. In particular, the existence of an abrupt transition from a stationary regime where oscillations are strongly damped to an oscillatory one with amplitudes of more than 10\% in the number of infected nodes predicted by the ODE model is clearly observed in the simulations over different types of networks. Moreover, when the size of the network increases ($N=10000$), periodic epidemics with amplitudes larger than 5\% carried out on regular random networks fall within the oscillatory regime in the parameter space predicted by the ODE model (see Fig.~\ref{fig:diagrama_hopf_N10000}). So, the robustness of the predictions about the existence of oscillations due to pure behavioural changes has been established.    

An interesting example of oscillating epidemics is given by the evolution of the incidence rate of sexual transmitted diseases (STDs) during the last decades. The reemergence of STDs like syphilis and gonorrhea occurring since the mid-1990s \cite{WCh} has been associated with a decrease in awareness after the introduction of the antiretroviral therapy for HIV and, indeed, it appeared after an incidence decline in the 1980s. This decline coincided with the emergence of the global AIDS pandemic and has been attributed to preventive behavioural changes in response to HIV campaigns during that time \cite{Fenton}. However, it has been also claimed that the drop in incidence before 1984 occurred too early to be ascribed to such induced behavioural changes and may be part of a long term cyclic trend of this type of diseases (although the nature of this periodic behaviour remains unexplained) \cite{Green}.

\section*{Acknowledgments}
D.J. and D.R. have been partially supported by the \emph{Agencia Estatal de Investigación} and \emph{Ministerio de Ciencia, Innovación y Universidades} grant MTM2017-86795-C3-1-P, and 
D.J. and J.S. have been partially supported by the grant PID2019-104437GB-I00 of the
\emph{Ministerio de Ciencia e Innovaci\'on} of the Spanish government. D.J. and D.R. are members of the \emph{Consolidated Research Group} 2017 SGR 1617 funded by the \emph{Generalitat de Catalunya}. J.S. is member of the \emph{Consolidated Research Group} 2017 SGR 01392 of the \emph{Generalitat de Catalunya}.

\bibliographystyle{apsrev4-2}
\bibliography{references}

\end{document}